\documentclass[preprint,amsmath,amssymb]{revtex4}

\usepackage{color}
\usepackage{graphicx}
\usepackage{dcolumn}
\usepackage{bm}
\input epsf

\def\u{{\rm \bf u}}

\def\x{{\rm \bf x}}

\newcommand{\ks}{\textcolor{black}} 

\newcommand\beq{\begin{equation}}
\newcommand\eeq{\end{equation}}
\begin{document}


\title{Motion of an air bubble under the action of thermocapillary and buoyancy forces}
\author{Manoj Kumar Tripathi and Kirti Chandra Sahu$^{\dagger}$}\email{ksahu@iith.ac.in}
\affiliation{Indian Institute of Science Education and Research Bhopal 462 066, Madhya Pradesh, India \\
$^{\dagger}$Department of Chemical Engineering, Indian Institute of Technology Hyderabad, Sangareddy 502 285, Telangana, India}

 \date{\today}
 
 \begin{abstract}
A novel way to handle surface tension gradient driven flows is developed in the volume-of-fluid (VoF) framework. Using an open source Navier-Stokes solver, {\it Basilisk}, and the present formulation, we investigate thermocapillary migration of drops/bubbles in a surrounding medium. Several validation exercises have been performed, which demonstrate that the present solver is a robust one to investigate interfacial flows with variable surface tension. It is well known that it is a challenging task to numerically model the tangential and normal surface forces arising due to interfacial tension. We have shown that the present method does not require the artificial smearing of surface tension about the interface, and thus predicts the theoretical value of the terminal velocity of bubble/drop migrating due to an imposed temperature gradient very well. It is also demonstrated that the present solver provides accurate results for problems exhibiting the gravity and thermocapillary forces simultaneously, and useful for systems with high viscosity and density ratios. 
\end{abstract}

\maketitle
\section{Introduction}
\label{sec:intro}
Thermocapillary flows are numerically very challenging {to model} due to the need of calculating the surface tension force along and across the interface separating the fluids. In gas-liquid systems, the difference of thermal conductivities of the phases results in large temperature gradient in both normal and tangential directions at the interface between the fluids. This in turn leads to the development of additional stresses along the interface. Using the formulation proposed by Brackbill {\it et al.} \cite{brackbill}, these stresses can be included into the Navier-Stokes equations as body forces, which act only at the interface, as
\begin{equation}
\rho \frac{D \u}{D t}  = \nabla \cdot \tau + \sigma \kappa {\bf n} \delta  (\x-\x_f) + \nabla_s \sigma \delta  (\x-\x_f)+\rho g \vec e_z. \label{model}
\end{equation}
Here $\rho$ represents density and $\sigma$ denotes the interfacial tension coefficient of the interface separating the fluids. $D \u/  D t$ is substantial acceleration and $\tau$ denotes the stress tensor. {$\delta (\x-\x_f)$ is a delta distribution function (denoted by $\delta$ hereafter) that is zero everywhere except at the interface, where $\x = \x_f$ is the position vector of a point at the interface.} $\kappa = \nabla \cdot {\bf n}$ is the curvature, ${\bf n}$ is the  unit normal to the interface pointing towards the outer fluid, $g$ is the acceleration due to gravity and $\vec e_z$ represents the unit vector in the vertically upward direction. The surface gradient operator is represented by $\nabla_s (\equiv \nabla - (\nabla \cdot {\bf n}){\bf n})$. In Eq. (\ref{model}), there are two terms associated with surface tension, namely, $\delta \sigma \kappa {\bf n}$ and $\delta \nabla_s \sigma$ which act normal and tangential to the interface, respectively. The later one $(\delta \nabla_s \sigma)$ is commonly known as Marangoni stress. This mechanism drives the flow in the vicinity of the interface and is always present in non-isothermal interfacial flows, and obviously can be important in a great variety of technological applications (see for instance Refs. \cite{Subramanian1992,Subramanian2002}). Apart from this, Marangoni stresses may develop in systems with bulk concentration gradients and surfactants.

In the present study, we consider a characteristic problem where thermal Marangoni stresses play a significant role, namely, the thermocapillary migration of drops and bubbles in a surrounding medium.  In such situations, the tangential (Marangoni) stresses drive the continuous phase towards the colder region, and the reaction of which helps to migrate the bubble/drop in the opposite direction. As most of the previous numerical studies isolate the Marangoni effect by considering the microgravity condition, we have validated our numerical solver by neglecting the effect of gravity. However, in several industrial applications gravity and thermocapillary stresses act simultaneously. The action of gravity/buoyancy force along with thermocapillary force may result in a more complex flow dynamics, as discussed below. Thus, we also study the migration of an air bubble in a liquid medium under the action of both buoyancy and thermocapillary forces by conducting three-dimensional (3D) numerical simulations. A schematic diagram of the problem considered is shown in Fig. \ref{geom}, wherein fluid `$A$' and fluid `$B$' designate the {continuous and the dispersed} phases, respectively.

A brief review of the previous studies on migration of bubbles/drops due to the thermocapillary forces is presented below. The thermocapillary migration of a bubble in a viscous fluid heated from below was first reported in the pioneering work of Young {\it et al.} \cite{Young1959}. In the system considered, there is a competition between the buoyancy (acting in the upward direction) and surface tension force (acting in the downward direction). By conducting experiments, they demonstrated that small bubbles move in the downward direction, whereas bigger bubbles moves in the upward direction. Thus, the thermocapillary force wins in case of small bubbles, but buoyancy overcomes the effect of thermocapillarity in case of big bubbles. They also derived an analytical expression of the terminal velocity of thermally driven migration of a spherical bubble in the microgravity condition. Later, Balasubramaniam \& Chai \cite{Balasubramaniam1987} extended the analytical solution to bubbles with small deformation from a spherical shape. By conducting an asymptotic analysis in the limiting case of large Reynolds and Marangoni numbers, Balasubramaniam \cite{Balasubramaniam1998} showed that the steady migration velocity, at leading order, is a linear combination of the velocity for purely thermocapillary motion and the buoyancy-driven rising velocity. 

For small Marangoni numbers, Zhang {\it et al.} \cite{Zhang2001} showed via a theoretical analysis that inclusion of inertia is crucial in the development of an asymptotic solution for the temperature field. Recently, Herrmann {\it et al.} \cite{Herrmann2008} and Brady {\it et al.} \cite{Brady2011} conducted numerical simulations of a droplet inside a rectangular box in the limit of zero Marangoni number (i.e. assuming the thermal conductivities of the fluids to be infinity) and for finite values of Marangoni number, respectively. They also neglect gravity/buoyancy in their numerical simulations. They showed that for low Marangoni numbers the drop rapidly settles to a quasi-steady state, whereas for high Marangoni numbers the initial conditions significantly affect the behaviour of the droplet. They compared the terminal velocity of the drop obtained from their numerical simulations with the theoretical prediction of Young {\it et al.} \cite{Young1959}. Welch \cite{Welch1998} demonstrated that for higher capillary numbers bubble deformation becomes important and the bubble continues to deform at later times, failing to reach a steady state. Herrmann {\it et al.} \cite{Herrmann2008} and Wu \& Hu \cite{Wu2012,Wu2013} also {reached to} the same conclusion for {the case of} large Marangoni numbers. Liu {\it et al.} \cite{liu2012} investigated thermocapillary migration of a bubble at high Marangoni numbers using a lattice Boltzmann method and showed that the terminal velocity of the bubble decreases with increasing Marangoni number.

Keh {\it et al.} \cite{Keh2002} numerically studied the motion of a spherical drop between two parallel plane walls and found that the droplet migration speed can be controlled by varying the thermal conductivity of the droplet and changing the imposed boundary conditions at the walls. Chen {\it et al.} \cite{Chen1991} found that inside an insulated tube with an imposed axial temperature gradient, which in turn develops the hydrodynamic retarding forces, the thermocapillary migration velocity of a spherical drop is always less than that in an infinite medium. This work was extended by Mahesri {\it et al.} \cite{Mahesri2014} to take into account the effect of interfacial deformation. 

In the recent times, an open-source code, {\it Gerris} \cite{popinet2009} has been used by several researches including our research group (see for instance \cite{tripathiNcomms2015}) to study interfacial flows. However, {this code does not have a module to handle surface tension gradients}. Seric {\it et al.} \cite{seric2017} {improved upon} this code to incorporate the tangential surface tension force term in the {\it Gerris} flow solver and {showed few validation studies for} thermocapillary migration of a droplet in systems without gravity. To the best of our knowledge, this is the only computational study which {aimed at accurately modeling the tangential surface tension forces using a height-function like approach in the VoF framework}, albeit for microgravity systems, to investigate migration of a bubble/drop in non-isothermal systems.


Few researchers (see for instance Ref. \cite{Merritt1993}) theoretically considered the migration of a bubble under the influence of buoyancy and thermocapillary forces. Merritt {\it et al.} \cite{Merritt1993} demonstrated by plotting streamline patterns that the system can exhibit complex flow structures and the intuition developed from the gravity driven migration is not good enough for the bubbles/drops which experience both buoyancy and thermocapillary forces simultaneously. Recently, Tripathi {\it et al.} \cite{tripathi2015} conducted axisymmetric simulations by considering a quadratic dependence of surface tension on temperature (so called the `self-rewetting' fluids), and investigated buoyancy-driven rise of a bubble inside a tube imposing a constant temperature gradient along the wall. They found that for sufficiently large surface tension and moderate inertia, the bubble motion can be reversed and eventually the bubble can be arrested near the position of minimum surface tension. However, they neglected the {contribution of the surface gradient term} in their numerical simulations.  

In the present study, we have developed a robust numerical solver to handle Marangoni stresses and implemented this module in an open source code, {\it Basilisk}, developed by Popinet and co-workers \cite{popinet2009}. First, the solver has been validated extensively by comparing with the previous experimental, theoretical and computational studies. Then using this solver, we investigate the migration of an air bubble/drop in another medium under the action of both buoyancy and thermocapillary forces. The later one is also associated with high density and viscosity contrasts, which are also known to be difficult to handle numerically.

The rest of the paper is organized as follows. A general description of the problem considered in given in Section \ref{sec:formulation}. The current numerical method is described in Section \ref{sec:num}. The results are presented in Section \ref{sec:dis}, wherein several validation exercises are also performed. Concluding remarks are given in Section \ref{sec:conc}.

\section{Formulation}
\label{sec:formulation}

\begin{figure}
\centering
\includegraphics[width=0.5\textwidth]{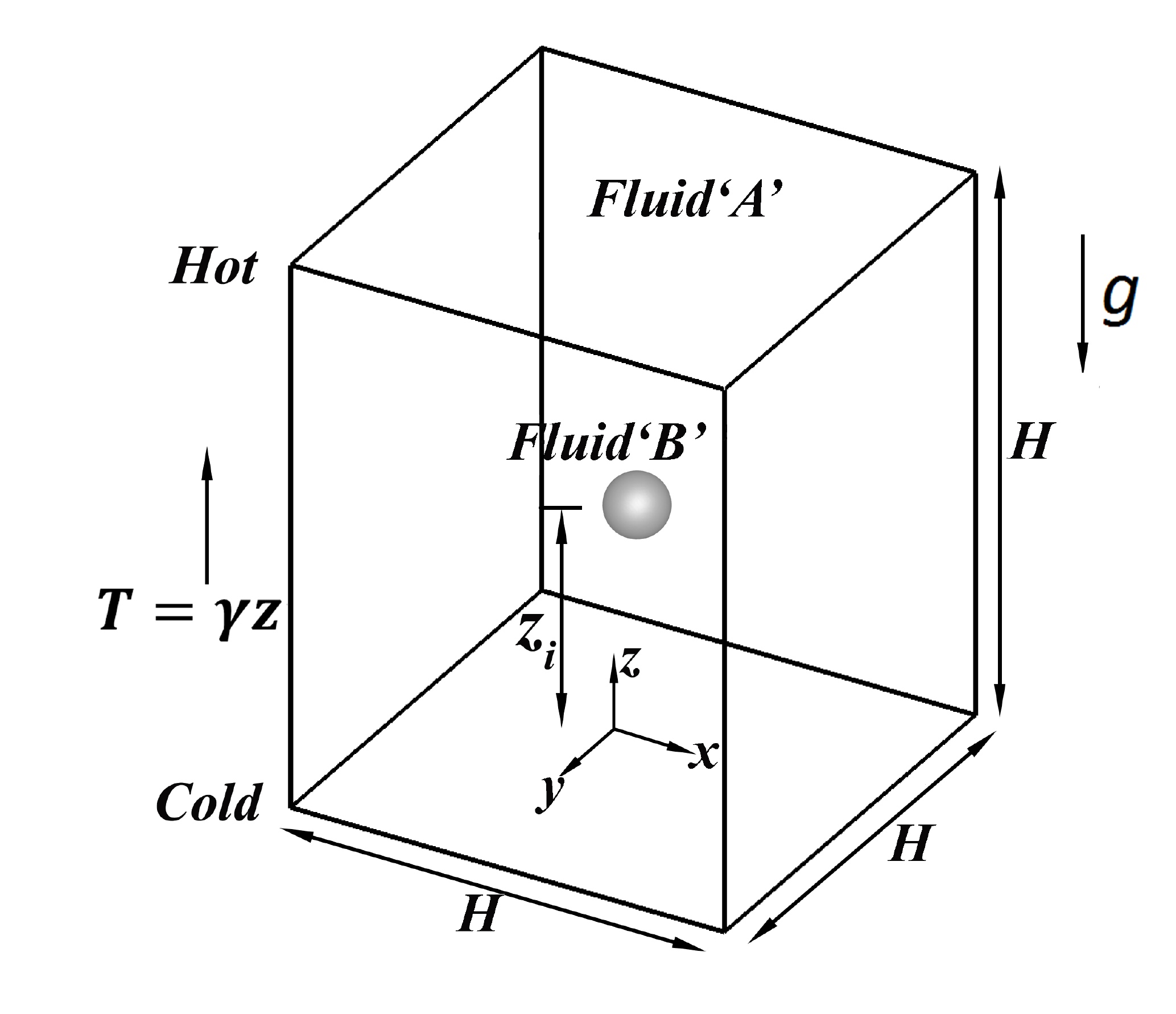} 
\caption{Schematic diagram showing the initial configuration of a bubble (fluid `$B$') of initial radius $R$ rising inside a cubical domain of dimension, $H$ containing a liquid (fluid `$A$') under the action of buoyancy. The bubble is placed at $z = z_i$ initially. The acceleration due to gravity, $g$ acts in the negative $z$ direction. A linear temperature variation $(T=\gamma z)$ is imposed at the side walls in the vertical direction with a constant gradient, $\gamma$.}
\label{geom}
\end{figure}


A general description of the characteristic problem considered in this study is given below. A schematic diagram is shown in Fig. \ref{geom}. Let's consider an initially spherical bubble of radius $R$ placed at the centre of a cubical domain of size $H$. The bubble (dispersed phase) and the surrounding medium (continuous phase) are designated by fluid `$B$' and `$A$', respectively. The acceleration due to gravity, $g$ acts in the negative $z$ direction, as shown in {Fig. \ref{geom}}.  A linear temperature variation is imposed at the walls in the vertical direction, given by $T= \gamma z$, where $\gamma$ is the temperature gradient in the $z$ direction. 

In order to perform the validation exercises, as considered by the previous studies, the gravity/buoyancy force is neglected in some of the cases. Also based on the previous investigations, either the dimensional or dimensionless forms of the governing equations are solved to compare our results with theirs.

The dimensional form of the governing equations are given by 
\begin{equation}
\nabla \cdot \u = 0,
\label{conti0}
\end{equation}
\begin{equation}
\rho \left [ \frac{\partial \u}{\partial t} + \u \cdot \nabla \u \right] = -\nabla p + \nabla \cdot \left [\mu (\nabla \u + \nabla \u^T) \right] + \delta  \sigma \kappa {\bf n}  +  \delta \nabla_s \sigma  + \rho g \vec e_z, \label{NS0}
\end{equation}
\begin{equation}
\rho c_p \left[{\partial T \over \partial t} + \u \cdot \nabla T \right ]= \nabla \cdot (\kappa \nabla  T), \label{temp0}
\end{equation}
where $\rho$, $\mu$, $c_p$ and $\kappa$ denote the density, viscosity, specific heat at constant pressure and thermal conductivity, respectively; ${\bf u}$, $p$ and $T$ denote the velocity, pressure, and temperature fields of the fluid, respectively, $t$ represents time; $\kappa = \nabla \cdot {\bf n}$ is the curvature, ${\bf n}$ is the  unit normal to the interface {pointing towards fluid `$A$'}; $\sigma$ represents the interfacial tension coefficient of the interface separating fluids `$A$' and `$B$'. A Cartesian co-ordinate system $(x,y,z)$ is used. The components of the velocity vector ${\bf u}$ are represented by $u$, $v$ and $w$ in the $x$, $y$ and $z$ directions, respectively. 

The following advection equation for the volume fraction of the liquid phase, $c$, is solved using a volume-of-fluid framework in order to track the interface separating the fluids: 
\begin{equation}
\frac{\partial c}{\partial t}+{\bf u}\cdot \nabla c = 0.
\label{adv0}
\end{equation}

The viscosity dependence on the temperature and the volume fraction of the liquid phase, $c$, which takes on values between 0 and 1 for the air and liquid phases, respectively, is given by \cite{nahme,tripathi2015}:
\begin{equation}
{\mu}=  c \mu_A e^ {-\left ( {T -T_{ref}\over T_{ref}} \right)} + (1-c) \mu_B  \left \{ 1+  \left ( {T - T_{ref} \over T_{ref}} \right)^{3/2} \right \}, 
\end{equation}
where $\mu_A$ and $\mu_B$ are the viscosity of the liquid and air phases at the reference temperature temperature, $T_{ref}$. 

The density, thermal conductivity and specific heat at constant pressure are assumed to be constants for each phase, which are given by
\begin{eqnarray}
\rho = \rho_A c + \rho_B (1-c), \\
\kappa = \kappa_A c + \kappa_B (1-c), \\
c_p = {c_p}_A c + {c_p}_B (1-c),
\end{eqnarray}
respectively. Here, $\rho_A$ and $\rho_B$ denote the density, $\kappa_A$ and $\kappa_B$ represent the thermal conductivity and ${c_p}_A$ and ${c_p}_B$ represent the specific heat at constant pressure of the fluid `$A$' and fluid `$B$', respectively. 

The following functional dependence of the surface tension on temperature is used:
\begin{equation}
\sigma = \sigma_0 -\beta  (T-T_{ref}),
\label{eq:sigma}
\end{equation}
where $\beta \equiv - {d \sigma \over d T} \left. \right|_{T_{ref}}$.

\subsection{Scaling}
\label{dimless}
We employ the following scaling in order to render the governing equations dimensionless:
$$
(x,y,z,z_i,z_m) ={R} \left({\widetilde x, \widetilde y, \widetilde z, \widetilde z_i, \widetilde z_m}\right), \hspace{1mm} t=t_s \widetilde t, \hspace{1mm} {\bf u}= V \widetilde{\bf u}, \hspace{1mm} p= \rho_A {V^2} \widetilde p,
$$
\[
\hspace{1mm} \mu = \mu_{A}\widetilde \mu, \hspace{1mm} \rho = \rho_{A}\widetilde \rho, \hspace{1mm} \kappa =\kappa_{A} \widetilde \alpha,  \hspace{1mm} c_p ={c_p}_{A} \widetilde c_p, \hspace{1mm} T = \widetilde T T_{ref} + T_{ref}, 
\]
\begin{equation}
\sigma=\sigma_0\widetilde{\sigma}, \hspace{1mm} \beta=\frac{\sigma_0}{T_{ref}}M, \hspace{1mm} \gamma = { \Gamma T_{ref}  \over R},
\label{eq:scaling}
\end{equation}
where tildes designate dimensionless quantities and $\sigma_0$ is the surface tension at the reference temperature, $T_{ref}$. The velocity scale, $V$ is ${\beta \gamma R / \mu_A}$ and the time scale, $t_s$ is $\mu_A / \beta \gamma$. Here, $M$ and $\Gamma$ represent the dimensionless $\beta$ and imposed temperature gradient at the side walls in the $z$ direction, respectively.

The governing dimensionless equations (after dropping tilde notations) are given by 
\begin{equation}
\nabla \cdot \u = 0,
\label{conti}
\end{equation}
\begin{eqnarray}
\rho \left[ {\partial \u \over \partial t} + \u \cdot \nabla \u \right]= -\nabla p + {1 \over Re} \nabla \cdot \left [\mu (\nabla \u + \nabla \u^T) \right] -  {\rho \over Fr}{\bf j} \nonumber \\ + {\delta \over Re Ca} \left [\kappa(1-{M} T ){\bf n} - M \nabla_s T \right] , \label{NS}
\end{eqnarray}
\begin{equation}
{\partial T \over \partial t} + \u \cdot \nabla T = {1 \over Ma} \nabla \cdot ( \alpha \nabla  T),
\label{temp}
\end{equation}
\begin{equation}
{\partial c \over \partial t} + \u \cdot \nabla c = 0,
\label{adv1}
\end{equation}
where ${Re} \equiv \rho_A V R/\mu_A$ denotes the Reynolds number, $Ma \equiv  {V R \rho_A {c_p}_A / \kappa_A}(\equiv Re Pr)$ is the Marangoni number, $Pr \equiv {\mu_A  {c_p}_A/ \kappa_A}$ is the Prandtl number, $Ca \equiv V \mu_A /\sigma_0$ is the capillary number and $Fr \equiv {V^2 / g R}$ is the Froude number. 

The dimensionless viscosity, $\mu$ is given by:
\begin{equation}
{\mu}=  c e^{-T} + (1-c) \mu_r \left ( 1+ T^{3/2} \right),
\label{vis:model}
\end{equation}
where $\mu_r \equiv {\mu_B / \mu_A}$ is the viscosity ratio. The dimensionless density, $\rho$ and thermal diffusivity, $\alpha \left( \equiv \kappa/\rho c_p \right) $ are given by:
\begin{eqnarray}
\rho = c + \rho_r (1-c),  \\
\alpha =  c + \alpha_r (1-c),
\end{eqnarray}
respectively, wherein $\rho_r \equiv {\rho_B/ \rho_A}$ and $\alpha_r \equiv {\alpha_B/ \alpha_A}$.

\section{Numerical method}
\label{sec:num}

A Navier-Stokes solver with VoF interface tracking algorithm, {\it Basilisk} \cite{popinet2009}, has been chosen as a starting point for the implementation of the presented method. The VoF advection algorithm employed is non-diffusive and conservative in nature \cite{weymouth2010}. Moreover, the calculation of surface tension force is balanced by pressure gradient exactly to the machine accuracy along with a height-function based interface curvature estimation. Although the code allows adaptive refinement of the mesh, we restrict the adaptive refinement to a region a few cells away from the interface; however, the extension to variable grid sizes throughout the domain is straightforward \cite{popinet2003}. 

The proposed method of incorporating the Marangoni forces in the Navier-Stokes equations employs a method similar to the one used for computing the curvature using height-functions. As shown in Eq. (\ref{NS0}), the Marangoni force term is the surface gradient of the surface tension coefficient. Computation of this term poses several difficulties for interface capturing techniques, in contrast to the interface tracking techniques. Identifying the exact values of the surface tension coefficient at the interface is generally done by some kind of averaging of the values in the cells surrounding the interfacial cell. An averaging technique similar to the computation of height-functions improves the accuracy significantly. This has recently been implemented in an open-source code, {\it Gerris} \cite{popinet2009}, by Seric {\it et al.} \cite{seric2017}. The authors have presented and compared the results for systems without gravity. To the best of our knowledge, our results are more promising and have much less error when validated against theoretical results as compared to the other numerical codes reported so far. 

In the following text, the numerical approximation of the surface gradient of the coefficient of surface tension will be discussed. In most of the interface capturing techniques, the following identity is used to express the gradient,   
\begin{equation}
\nabla_s \sigma = \nabla \sigma - {\bf n}({\bf n} \cdot \nabla \sigma).
\end{equation} 

Although many researchers have used this formula with caution that the surface tension coefficient is not defined on the either side of the interface, their results do not match exactly with the theoretical predictions. Also, this transformation cannot be used when the gradient in surface tension is due to surfactants. An alternative to calculating the gradients as a sum of Cartesian components is to compute the surface gradient directly and subsequently derive the components in the Cartesian directions from this value. This has been demonstrated in the work by Seric {\it et al.} \cite{seric2017} in considerable detail. Following this work, the Marangoni force per unit volume can be written as:
\begin{equation}
f_s = \left( {\partial \sigma \over \partial s_1} {\bf t_1} + {\partial \sigma \over \partial s_2} {\bf t_2} \right) \delta,
\label{surfgrad}
\end{equation}
where, $s_1$ and $s_2$ are the coordinates in the plane tangential to the interface, and ${\bf t_1}$ and ${\bf t_2}$ are the unit vectors in the corresponding coordinate directions. The force is calculated only at the interface. In this approach, we first compute an average of surface tension coefficient on the interfacial cells. Thereafter, we find the numerical approximation for the partial derivatives in Eq. (\ref{surfgrad}), which is then transformed to the components in the Cartesian coordinates using the geometrical information of the phase interface. Most of these steps are similar to those mentioned in Seric {\it et al.} \cite{seric2017}, with some simplifications and improvements to the gradient calculations. Therefore, the details will be given only for the new contributions from our side.

To start with, an auxiliary surface tension field, $\sigma_c$ is defined for each row or coloumn (hereafter, rows will also be known as coloumns in $x$, $y$ or $z$ directions) of cells such that $\sigma_c$ is a volume weighted average in that coloumn. For instance, for a coloumn in $x$-direction,
\begin{equation}
\sigma_{cx} = {\sum_{i} C_{i,j,k}\sigma_{i,j,k} \over \sum_{i} C_{i,j,k}}.
\label{sigcx}
\end{equation}
wherein, $C_{i,j,k}$ and $\sigma_{i,j,k}$ are the discretized form of the volume fraction field $c$ and the surface tension field $\sigma$ in a computational cell with the indices $(i,j,k)$. It should be noted that the right hand side of Eq. (\ref{sigcx}) will only be calculated for coloumns which contain the interface, and the right hand side will be equal to $\sigma$ if only one cell is cut by the interface in the coloumn under consideration. 
In the case of two spatial dimensions, there is only one unit vector required to form the basis for all tangential vectors to the interface, i.e., ${\bf t_1}$. For each interfacial cell, there are two possible auxiliary surface tension fields, obtained from coloumns in $x$ and $y$ directions. 
In this case, the surface gradient is approximated by a derivative of $\sigma_{cx}$ when the $x$-component of the normal vector to the interface is greater than the $y$-component, otherwise $\sigma_{cy}$ is used. This is similar to the choice exercised in the computation of curvature using height-functions defined in coloumns oriented in different Cartesian directions \cite{popinet2009}. For a derivative of $\sigma_{cx}$, for instance, the following difference approximation is employed,
\begin{equation}
\left( {\partial \sigma_{cx} \over \partial y} \right)_{i,j} = {\sigma_{cx, j+1} - \sigma_{cx, j-1} \over \Delta s}.
\end{equation}
The length of the interface between the two coloumns $(j-1)$ and $(j+1)$ is calculated using the geometry of the interface (Basilisk solver; {\it http://basilisk.fr/src/geometry.h}). The tangent vector ${\bf t_1}$ is chosen such that it is perpendicular to the interface normal and its direction is towards increasing $y$-coordinate to account for the sign of $\Delta s$. Similar procedure is applicable to an interface segment having the $y$-component of its normal vector greater than the $x$-component.

\begin{figure}
\centering
(a) \hspace{3.8cm} (b) \\
\includegraphics[width=0.5\textwidth]{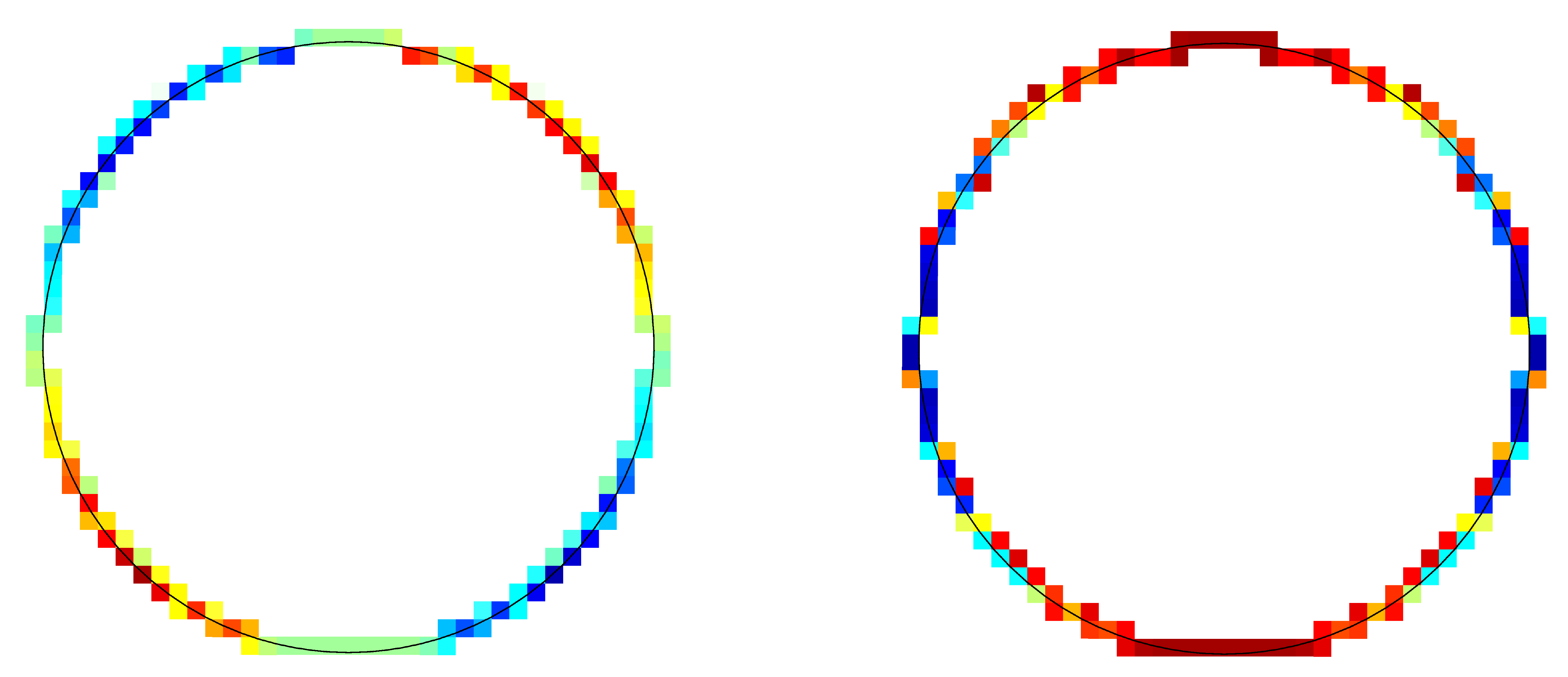} 
\caption{The components of surface tension gradients (a) $f_{sx}$ and (b) $f_{sy}$ at the interface in a two-dimensional system. In panel (a), green color represents zero, and red and blue represent the maximum (positive) and minimum (negative) values, respectively. In panel (b), red color represents zero, and blue represents the minimum (negative) value. The parameter values considered to plot these diagrams correspond to the case shown later in Fig. \ref{hermann}.}
\label{2Dsf}
\end{figure}

For 3D flows, there are two bases for a tangent vector to the interface, ${\bf t_1}$ and ${\bf t_2}$. We follow the discussion in Seric {\it et al.} \cite{seric2017} to fix the relative directions of the two bases vectors as follows. If the greatest component of the normal to the interface is in $x$-direction, the tangent vector components we consider are:
$ {\bf t_1} = (t_{1x}, 0, t_{1z}),$ and 
$ {\bf t_2} = (t_{2x}, t_{2y}, 0).$
 
For the coloumns in $x$-direction, the components of forces in the Cartesian directions can be found as
\begin{eqnarray}
f_{sx} &=& \left({\partial \sigma_{cx} \over \partial s_{1x}} t_{1x} + {\partial \sigma_{cx} \over \partial s_{2x}} t_{2x} \right)\delta, \\
f_{sy} &=& {\partial \sigma_{cx} \over \partial s_{2y}} t_{2y}\delta, \\
f_{sz} &=& {\partial \sigma_{cx} \over \partial s_{1z}} t_{1z}\delta.
\end{eqnarray} 
Some typical plots for surface force components, $f_{sx}$ and $f_{sy}$ at the interface for a two-dimensional (2D) system is shown in Fig. \ref{2Dsf}(a) and (b). This case is shown to demonstrate that the surface forces are calculated exactly at the interfacial cells. It can be seen that the surface tension decreases as we move in the positive $y$ direction. A similar procedure is also followed for 3D systems. We compute the delta distribution function ($\delta$) from the length of the interface per unit area in 2D simulations, and area of the interface per unit volume in 3D simulations. This is one of the differences between our algorithm and the algorithm used by Seric {\it et. al} \cite{seric2017}.

\section{Results and discussion}
\label{sec:dis}

\subsection{Bubble rise in an isothermal condition: comparison with Bhaga \& Weber \cite{bhaga1981}}
We have started our validation exercises by comparing the dynamics of an air bubble rising in aqueous sugar solutions of differing concentrations in an isothermal condition, as studied experimentally by Bhaga \& Weber \cite{bhaga1981}. Based on the parameters used to generate Figure 3 of Bhaga \& Weber \cite{bhaga1981}, the viscosity $(\mu_r)$ and density $(\rho_r)$ ratios are fixed at $8.153 \times 10^{-6}$ and $7.473 \times 10^{-4}$. A large computational cubic domain of size $H=120 R$ is considered for this study, such that the boundary effect can be neglected. Initially (at $t=0$), an air bubble is assumed to be stationary at $z_i=7R$ and starts to rise at time, $t>0$ due to the buoyancy force. Free-slip and no-penetration conditions are imposed at all the boundaries of the computational domain to mimic the unconfined boundaries. Wavelet error based adaptive mesh refinement has been used to increase the accuracy at the interface and the regions with higher velocity gradients (with respect to a tolerance value of $10^{-3}$). These regions are refined with approximately 68 cells per bubble diameter.

In the formulation presented in Section \ref{sec:formulation}, using $\sqrt{gR}$ as the velocity scale, instead of $\beta \gamma R / \mu_A$, as the system at hand is isothermal, we get the following dimensionless governing equations: 
\begin{equation}
\nabla \cdot \u = 0,
\label{conti}
\end{equation}
\begin{eqnarray}
\rho \left[ {\partial \u \over \partial t} + \u \cdot \nabla \u \right]= -\nabla p + {1 \over Ga} \nabla \cdot \left [\mu (\nabla \u + \nabla \u^T) \right] -  {\rho}{\bf j} + {\bf n} {\delta \over Eo} \nabla \cdot {\bf n}, \label{NS3}
\end{eqnarray}
where ${Ga} \equiv \rho_A g^{1/2} R^{3/2}/\mu_A$ denotes the Galilei number and $Eo \equiv {\rho_A r R^2/ \sigma_0}$ is the E\"{o}tv\"{o}s number. These equations are solved to compare the results obtained from the present simulations with those of Bhaga \& Weber \cite{bhaga1981}.

\begin{figure}[h]
\centering
\includegraphics[width=0.5\textwidth]{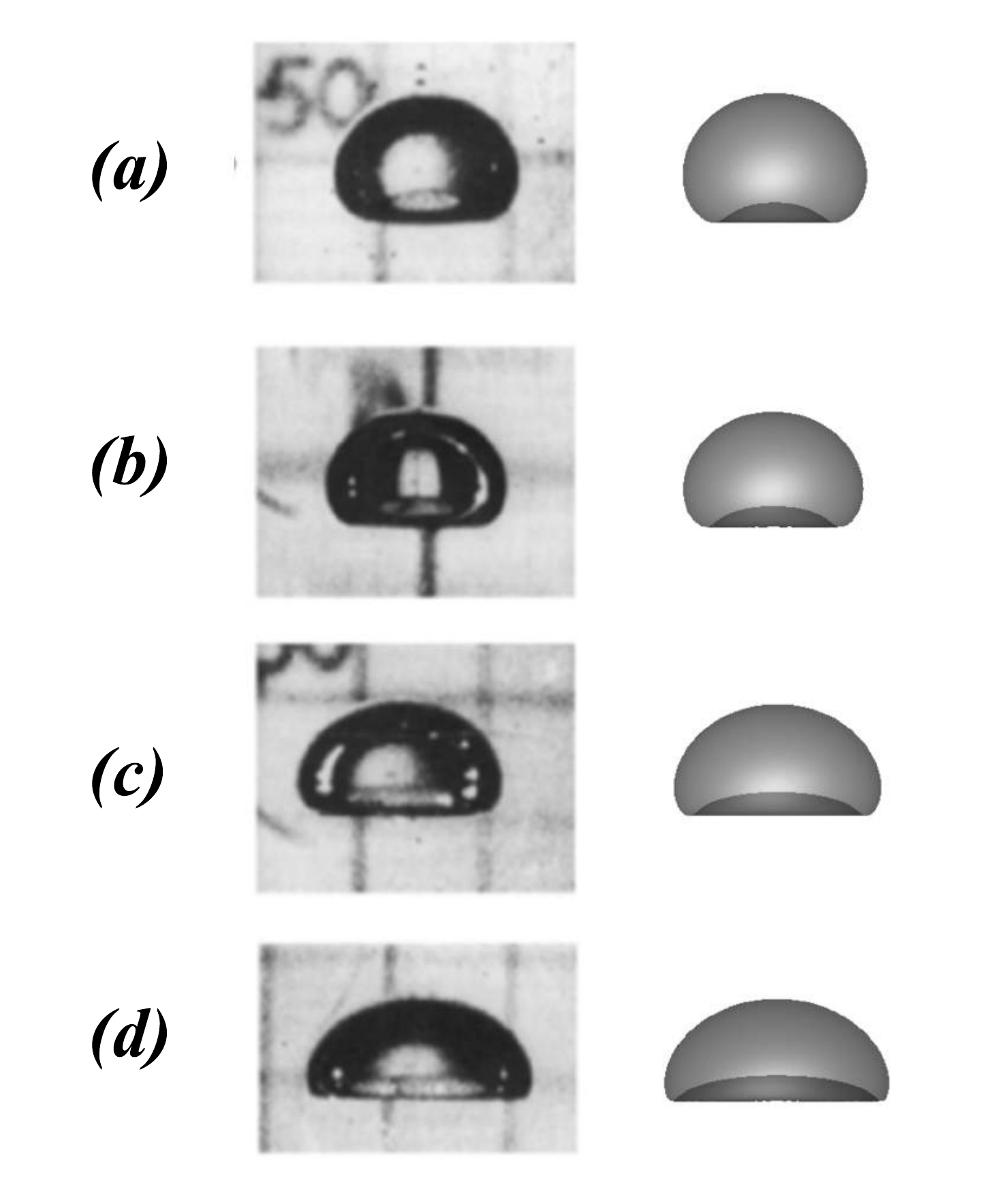} 
\caption{Comparison the terminal shapes of an air bubble rising in aqueous sugar solutions obtained from our 3D numerical simulations with those of Bhaga \& Weber \cite{bhaga1981}. (a) $Ga=2.315$, (b) $Ga=3.094$, (c) $Ga=4.935$ and (d) $Ga=8.157$. The rest of the parameters are $Eo=29$, $\mu_r=8.152 \times 10^{-6}$ and $\rho_r=7.473 \times 10^{-4}$.}
\label{bhaga1}
\end{figure}

Bhaga \& Weber \cite{bhaga1981} used $Eo_{BW} \equiv 4 g R^2 \rho_A/\sigma$ and $Mo_{BW} \equiv g \mu_A^4/\rho_A \sigma^3$ as E\"{o}tv\"{o}s number and Morton number to describe their flow systems. In the present study, the Morton number $(Mo_{BW})$ can also be defined as $Eo^3/Ga^4$. It is to be noted that the Morton number is a constant for a given liquid-gas system. A suitable transformation gives the following relationships: 
\begin{eqnarray}
Ga &=& \left ( {Eo_{BW}^3 \over 64 Mo_{BW}} \right)^{1/4} {\rm and} \\
Eo &=& {Eo_{BW} \over 4}.
\end{eqnarray}

\begin{figure}[h]
\centering
(a) \hspace{5.0cm} (b) \\
\includegraphics[width=0.25\textwidth]{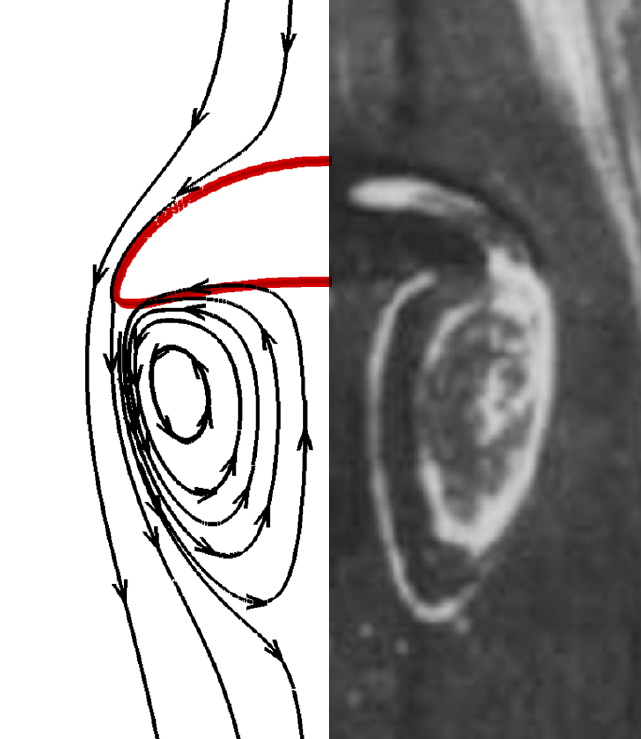} \hspace{10mm} \includegraphics[width=0.27\textwidth]{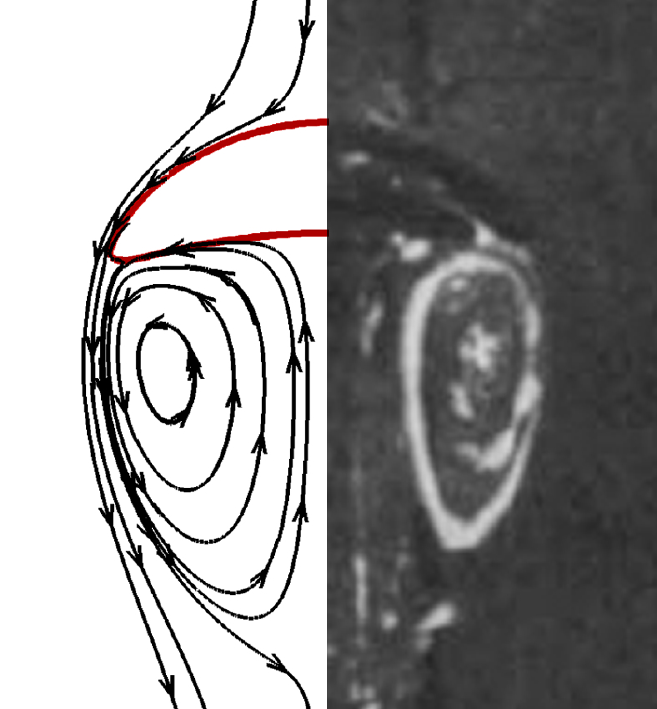} 
\caption{Comparison of streamline patterns at a cross-section along the axis of the domain when the bubble reaches the terminal shape obtained from the present simulations (on the left hand side of each panel) with those of Bhaga \& Weber \cite{bhaga1981} (right hand side of each panel). (a) $Ga=14.28$, $Eo=18.16$, and (b) $Ga=20.12$, $Eo=28.69$. The rest of the parameters are $\mu_r=8.152 \times 10^{-6}$ and $\rho_r=7.473 \times 10^{-4}$.}
\label{bhaga2}
\end{figure}

In Fig. \ref{bhaga1}, the terminal shapes of the bubble obtained from the present 3D numerical simulations for different $Ga$ values and $Eo=29$ are compared with the corresponding experimental results obtained by Bhaga \& Weber \cite{bhaga1981} (see their Figure 3). It can be seen that the shapes of the bubble obtained from our numerical simulations are in excellent agreement with those obtained experimentally \cite{bhaga1981}. As expected, the size of the dimple at the bottom of the bubble increases with increasing $Ga$ due to the increase in the strength of the wake region. The streamlines in the $x$-$z$ plane passing through the centre of gravity of the bubble are shown in Fig. \ref{bhaga2} for two sets of $Ga$ and $Eo$. The left and right hand sides of each panel present the results obtained from the numerical simulations and experiments, respectively. Here also, it can be observed that the numerically obtained streamlines patterns agree very well with the re-circulation zones observed in the experiments.


\subsection{Non-isothermal bubble rise in the limit of zero Marangoni number}

Next, we validate our solver for non-isothermal systems. The thermocapillary migration of a bubble is considered in the limit of zero Marangoni number. This corresponds to system with a fixed temperature (time-invariant) variation. The dimensional governing equations presented in Section \ref{sec:formulation} are solved to compare with the results of the previous studies. 

For this purpose, the migration of a bubble due to the presence of temperature gradient is considered, as theoretically studied by Young {\it et al.} \cite{Young1959} in the zero gravity condition. In the creeping flow regime (low Reynolds number), Young {\it et al.} \cite{Young1959} derived the terminal velocity of a neutrally buoyant spherical bubble of radius $R$, which is placed inside another infinitely unbounded fluid at rest. A time-invariant linear temperature gradient is imposed, which drives the bubble from the low temperature to the high temperature region. The imposed temperature profile implies that the thermal conductivity of the fluids is infinite, i.e $Ma=0$. In this condition, the terminal velocity of the bubble is given by \cite{Young1959} 
\begin{equation}
w_{YGB} = - {2 \beta \gamma R \over {6 \mu_A + 9 \mu_B}}.
\label{young}
\end{equation}

\begin{figure}[h]
\centering
\includegraphics[width=0.6\textwidth]{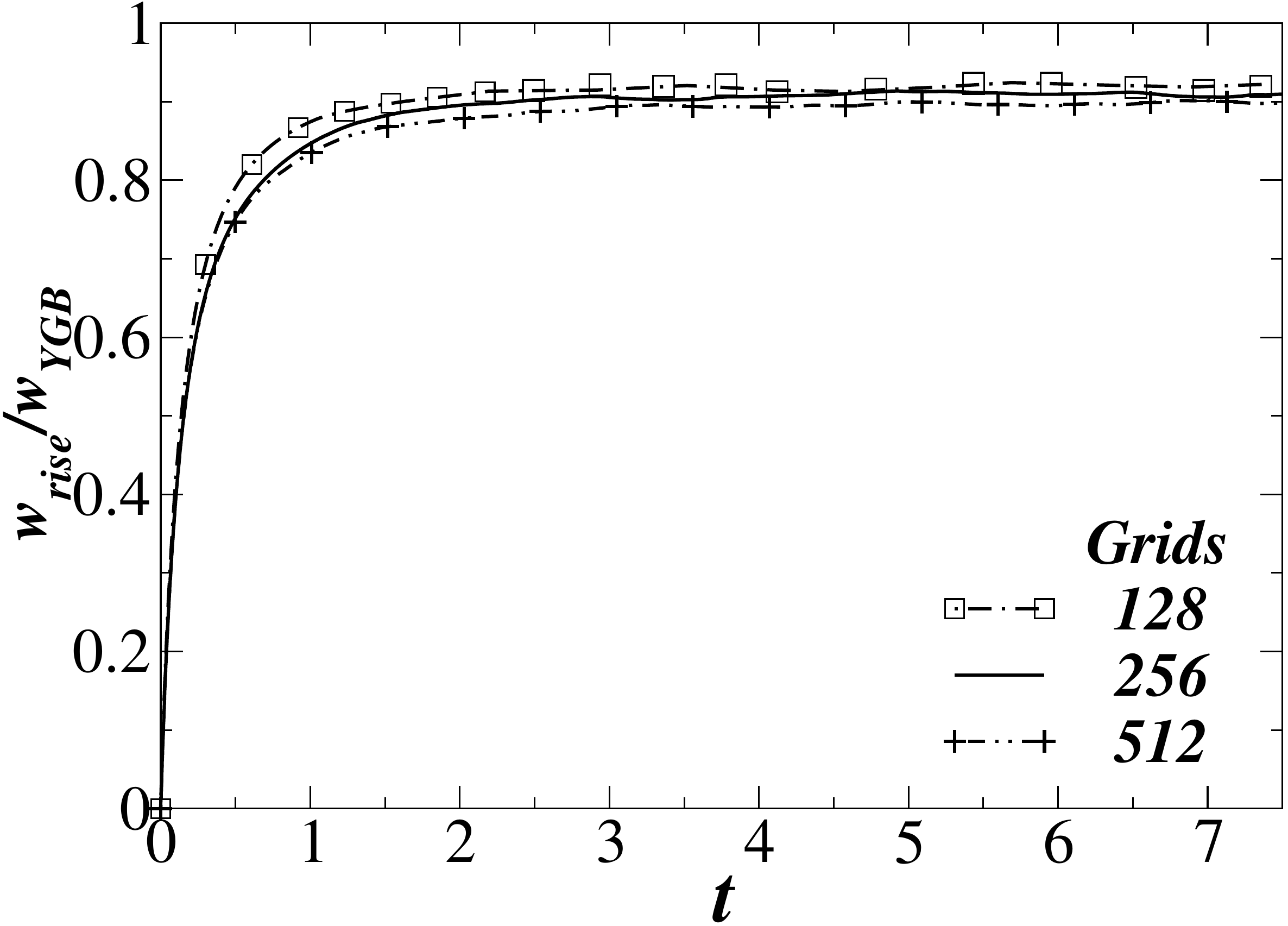}
\caption{Effect of grids on the normalized rise velocity of a bubble. The results obtained using three sets of grids (128, 256 and 512 grids in each direction of the 2D computational domain) are shown. The parameter values are the same as those used in Herrmann {\it et al.} \cite{Herrmann2008}.}
\label{convergence_test}
\end{figure}

\begin{figure}[h]
\centering
 \hspace{5mm}  (a) \hspace{7.5cm} (b) \\
\includegraphics[width=0.45\textwidth]{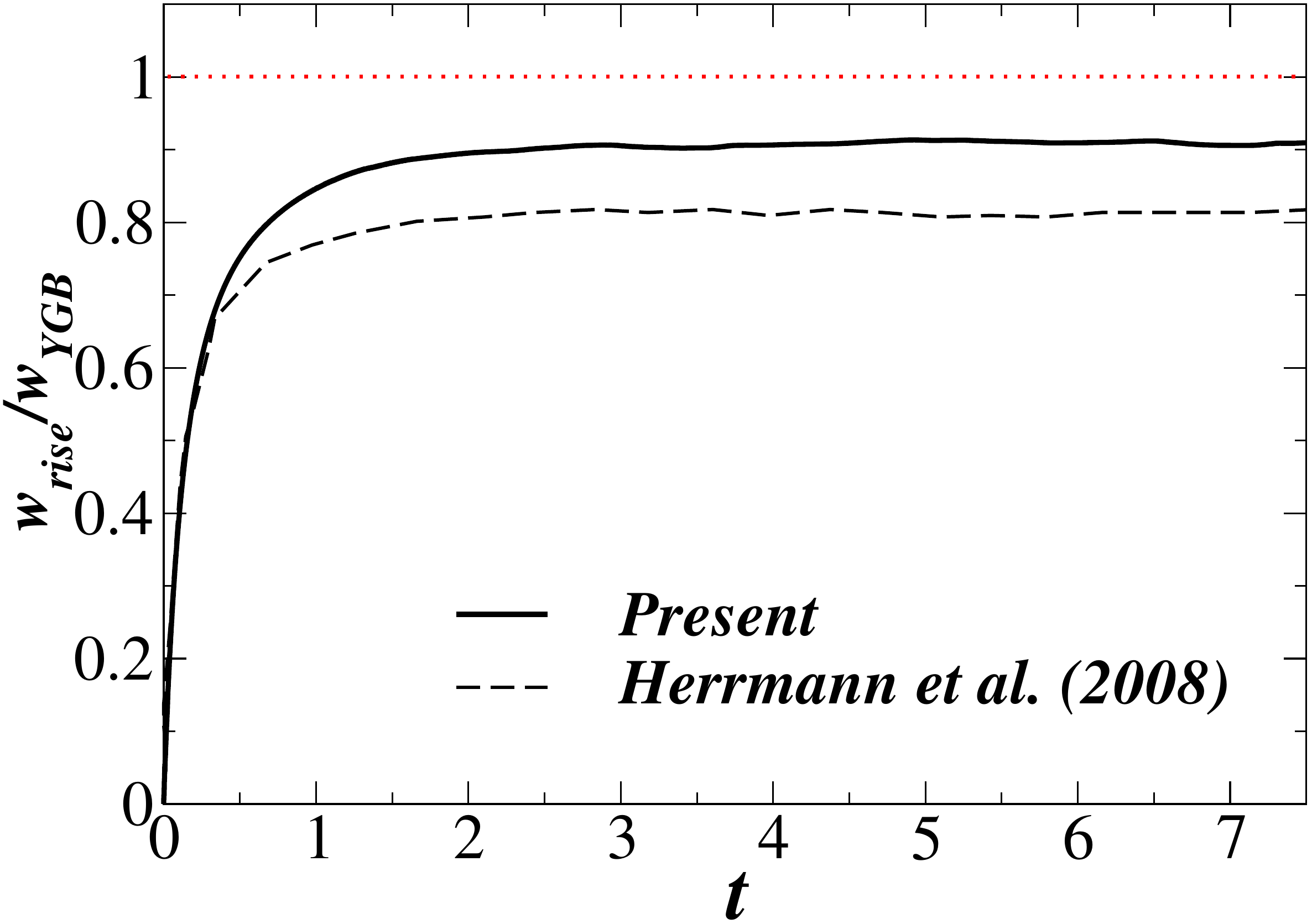} \hspace{5mm} \includegraphics[width=0.45\textwidth]{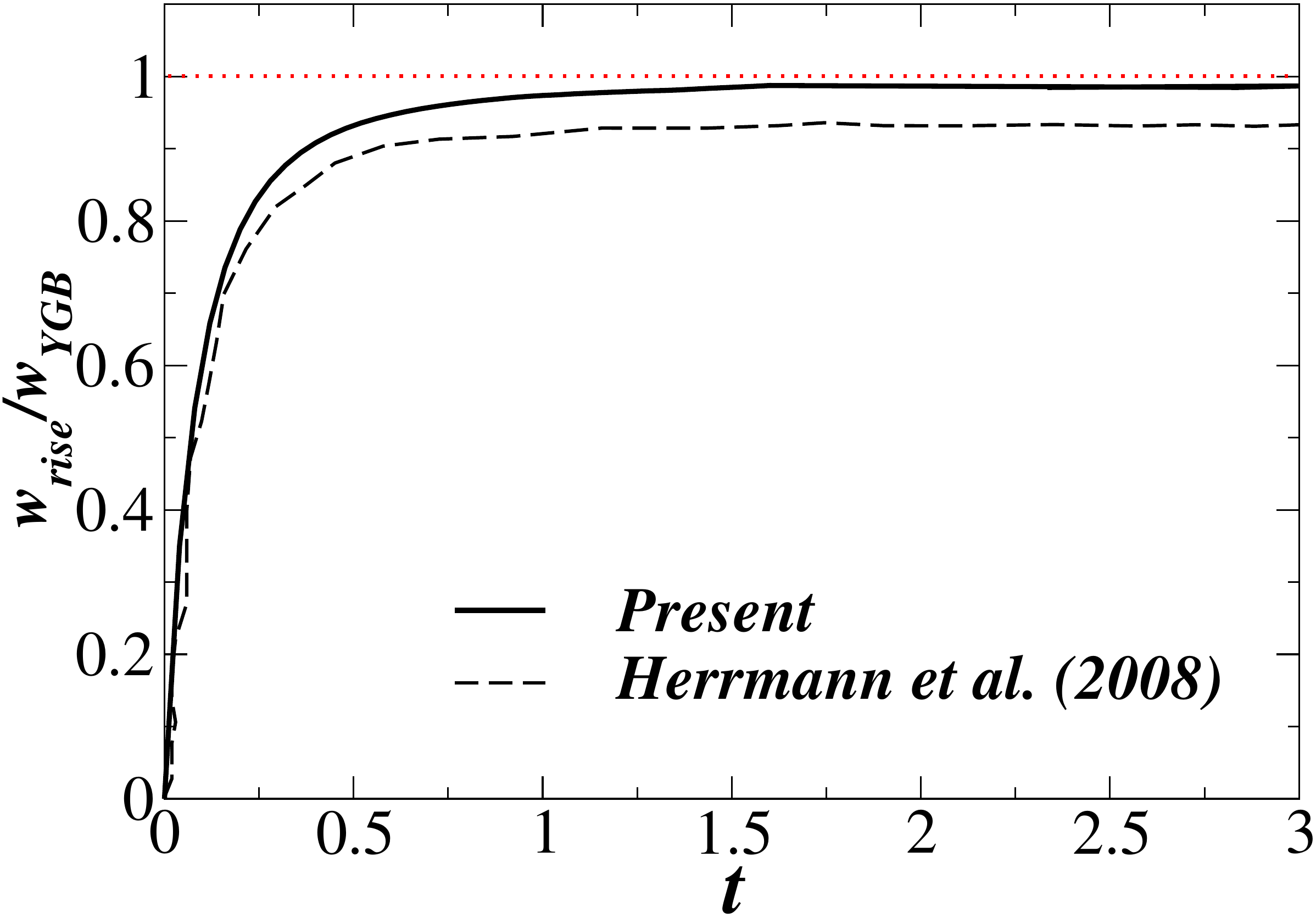} 
\caption{Comparison of the normalized rise velocity of a bubble  with Herrmann {\it et al.} \cite{Herrmann2008}. (a) 2D simulation, and (b) 3D simulation. The present simulations are conducted using 256 grids in each direction. The red dotted line in each panel represent the theoretical result of Young {\it et al.} \cite{Young1959}.}
\label{hermann}
\end{figure}

Before comparing the results obtained from our numerical simulations with the theoretical predication of Young {\it et al.} \cite{Young1959} and the previous computational results of Herrmann {\it et al.} \cite{Herrmann2008}, we have conducted a grid dependence test as shown in Fig. \ref{convergence_test}. A 2D computational domain of size $15R \times 15 R$ is considered, and the simulations are performed using 128, 256 and 512 grids (uniform) in each direction. It can be seen that they are in good agreement (the difference in the terminal velocity obtained using 256 and 512 grids is less than 1.5 \%).

Herrmann {\it et al.} \cite{Herrmann2008} conducted both 2D and 3D numerical simulations and compared their numerical results with the theoretical prediction of Young {\it et al.} \cite{Young1959}. Thus, we also considered a similar set-up as that of Herrmann {\it et al.} \cite{Herrmann2008}. The computation domain consists of a square box (in 2D) or a cubic (in 3D) of size $H=15 R$. The bubble of radius $R=0.5$ is placed at the centre of the computational domains. In the numerical simulations, no-slip and no-penetration boundary conditions are implemented at the top and bottom walls, and periodic boundary condition is used for all the side boundaries. A linear temperature field $T=\gamma z$ is imposed, such that $T=0$ at the bottom wall ($z=0$) and $T=1$ at the top wall ($z=15R$). Thus $\gamma \approx 0.133$. The other parameters considered in the numerical simulations are $\rho_A= \rho_B=0.2$, $\mu_A = \mu_B=0.1$, $\sigma_0=0.1$ and $\beta= -0.1$. The negative value of $\beta$ implies that the surface tension decreases with temperature (see Eq. (\ref{eq:sigma})). For this set of parameters, the theoretical rise velocity of the bubble, $w_{YGB} \approx 8.888 \times 10^{-3}$. After conducting a grid refinement test, uniform grids of $256^2$ and $256^3$ are used in the 2D and 3D simulations, respectively. The finest grids used by Herrmann {\it et al.} \cite{Herrmann2008} were also the same as the ones used in the present simulations.


\begin{figure}[h]
\centering
\includegraphics[width=0.5\textwidth]{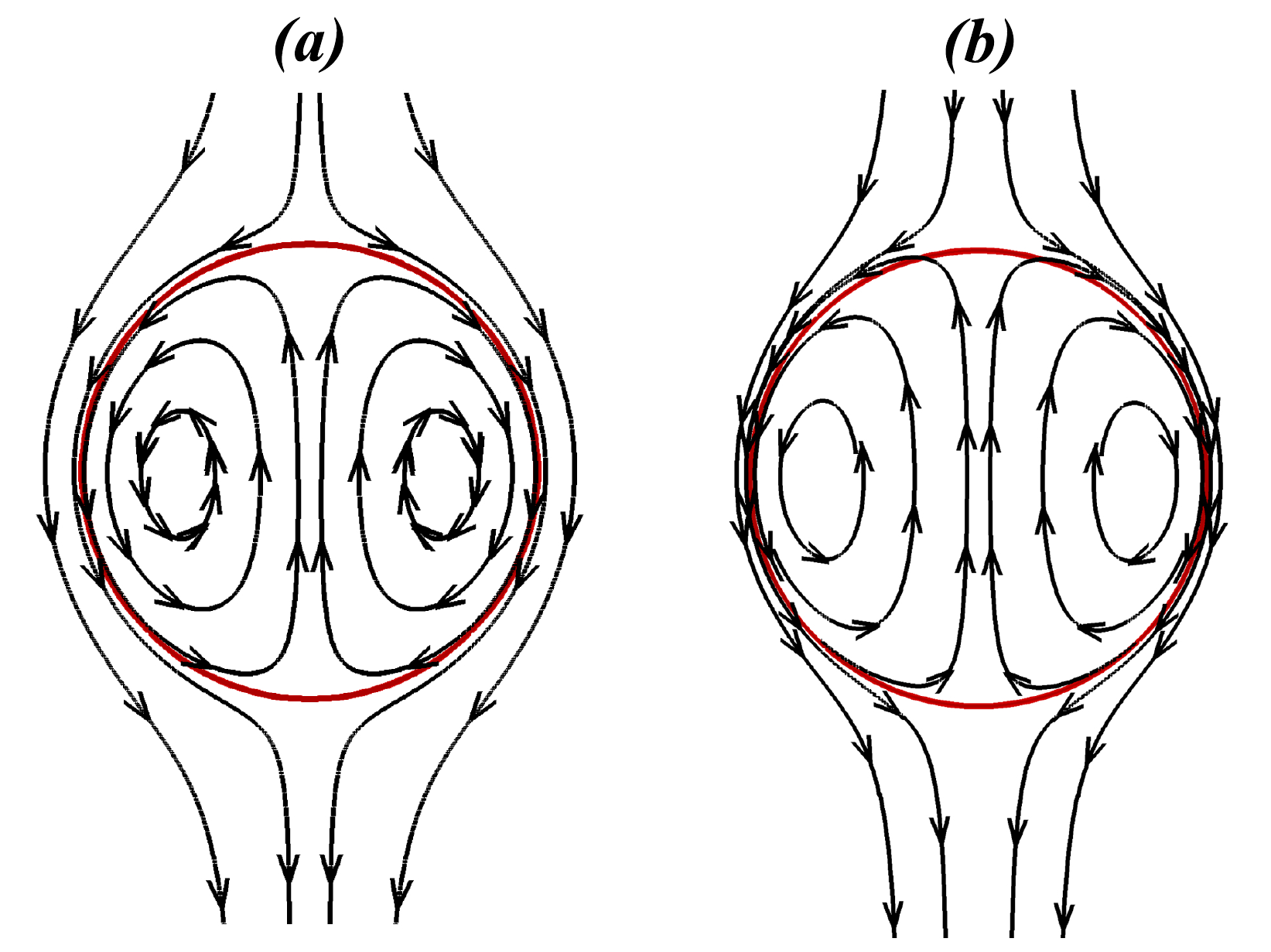}
\caption{Streamlines when the bubble reaches the terminal state obtained from the (a) 2D simulation, and (b) 3D simulation. The parameters are the same as those used to generate Fig. \ref{hermann}.}
\label{hermann2}
\end{figure}

The comparisons the rise velocity of the bubble $\left (w_{rise} \equiv { {\int_{V} c w d V } /  {\int_{V} c d V}} \right )$ normalised with the theoretical rise velocity of the bubble $(w_{YGB})$ versus time have been plotted in Fig. \ref{hermann}(a) (obtained from 2D simulation) and Fig. \ref{hermann}(b) (obtained from 3D simulation). The results of Herrmann {\it et al.} \cite{Herrmann2008} are shown by dotted lines, whereas our results are shown by solid lines in Fig. \ref{hermann}(a) and (b). In our 2D study (Fig. \ref{hermann}(a)), the percentage of error, which is defined as $(1 - w_{rise}/w_{YGB}) \times 100$, is 8.4 \%. This is much higher (as high as 18 \%) in the study of Herrmann {\it et al.} \cite{Herrmann2008}. The normalised rise velocity obtained in our 3D simulation (Fig. \ref{hermann}(b)) is very close to the theoretical result of Young {\it et al.} \cite{Young1959} (percentage of error is less than 1.6 \%), whereas it is 6.3 \% in the study of Herrmann {\it et al.} \cite{Herrmann2008}. For the same problem, by conducting simulation based on front-tracking method, Muradoglu \& Tryggvason \cite{Muradoglu08} reported the percentage of error to be 3 \%.  The streamlines patterns obtained from our two and 3D simulations at the terminal state are shown in Fig. \ref{hermann2}(a) and (b), respectively. The vortex and the flow patterns agree well with those obtained by Herrmann {\it et al.} \cite{Herrmann2008}. 

\begin{figure}[h]
\centering
\includegraphics[width=0.6\textwidth]{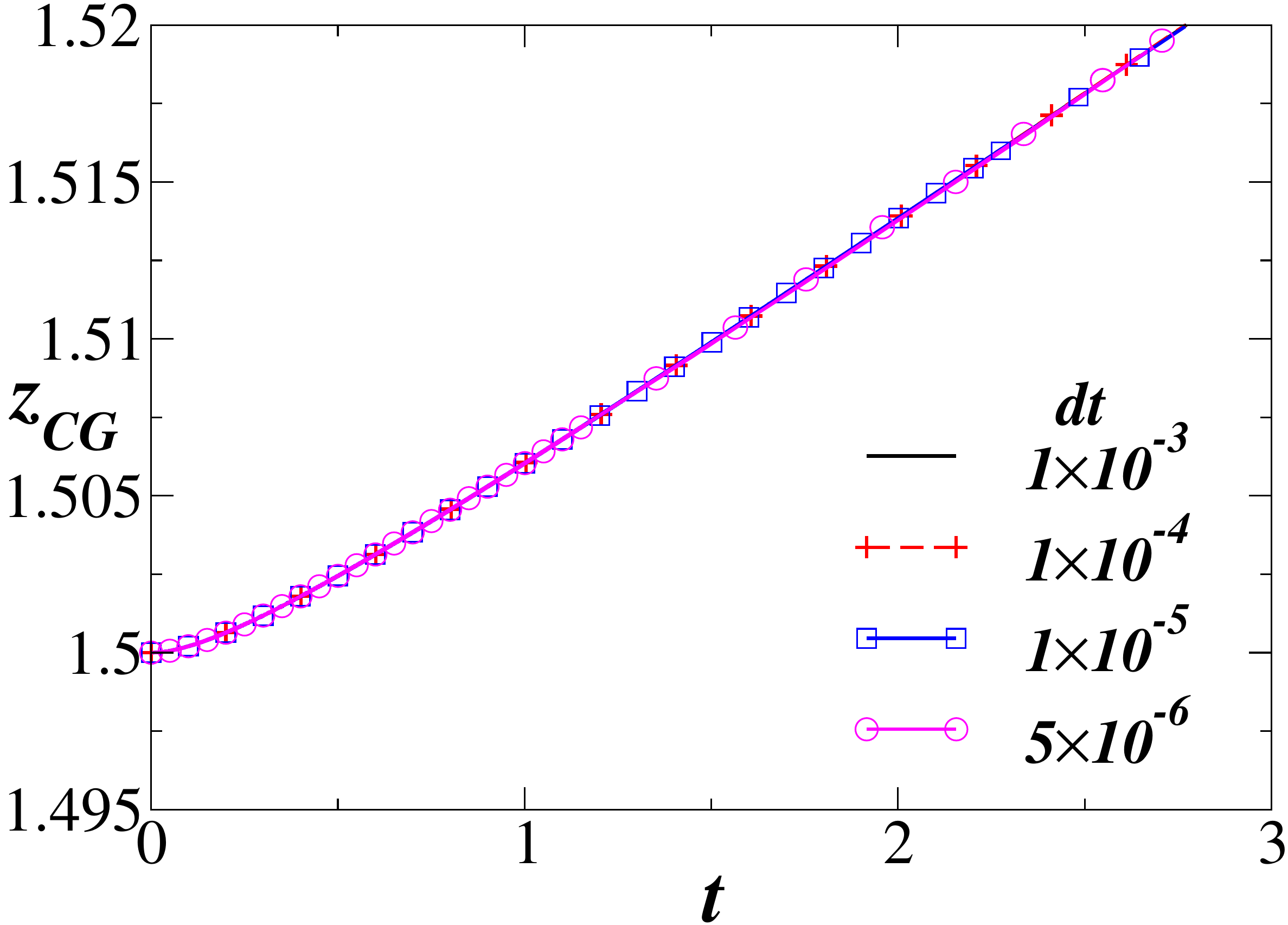}
\caption{Effect of time step, $dt$ on the variation of centre of gravity of the bubble with time. The parameter values considered are the same as those used to generate Fig. \ref{hermann}.}
\label{timestep}
\end{figure}

In order to show that larger time step can be used in the present solver than that used in the previous studies (Seric {\it et al.} \cite{seric2017}), we investigate the effect of time step on the rise dynamics in Fig. \ref{timestep}. The temporal variations of centre of gravity of the bubble obtained from 2D simulations using different time steps are shown in Fig. \ref{timestep}. The rise velocity presented in Fig. \ref{hermann} can be obtained by differentiating these results. It can be seen that the variations are indistinguishable for $dt < 1 \times 10^{-3}$. In contrast, Seric {\it et al.} \cite{seric2017} obtained converged results only for $dt < 10^{-5}$. We will discuss more about the comparison of our result with that of Seric {\it et al.} \cite{seric2017} in Section \ref{seric}.

\subsection{Non-isothermal bubble rise with finite Marangoni number} 

Next, we consider the migration of bubble in a non-isothermal system with finite Marangoni number. From dimensional governing equations (Eqs. (\ref{NS0})-(\ref{temp0})) and dimensionless governing equations (Eqs. (\ref{NS})-(\ref{temp})), it can be seen that there is a coupling between the Navier-Stokes and energy equations through the advection term in the energy equation. This coupling leads to interfacial Marangoni flow, which in turn reduce the tangential temperature gradient at the interface. 

\subsubsection{Comparison with Brady {\it et al.} \cite{Brady2011} } 
Three-dimensional simulations are performed by solving the dimensional set of governing equations (\ref{conti0})-(\ref{adv0}) without the buoyancy term in Eq. (\ref{NS0}) (reduced gravity condition). The problem is formulated in the same way as that of Brady {\it et al.} \cite{Brady2011}. The schematic diagram is shown in Fig. \ref{geom}, which is a cubic domain with $H=12R$. Initially a spherical bubble of radius $R=5.35$ mm is placed at $z=z_i=3R$ of the computational domain. The bubble moves due to the imposed temperature gradient, $\gamma$ at $t>0$. No-slip and no penetration boundary conditions are used at all the side walls and the Neumann boundary conditions for the velocity components are used at the top and bottom of the computational domain. A constant temperature ($T_0=283 K$) is maintained at the bottom of the computational domain ($z=0$) and a linear time-invariant temperature profile ($T=T_0 + \gamma z$) is imposed at all the side walls. The temperature at the top of the computational domain is also fixed at $T_1=T_0 + \gamma 12 R$. The temperature of the bubble is fixed at a temperature equals to the bulk fluid temperature at $z=3R$. A constant value of surface tension, $\sigma = 0.007$ N/m is used. Like in case of Brady {\it et al.} \cite{Brady2011}, Fluorinert FC-75 and silicone oil are used as the dispersed (bubble) and continuous (surrounding) fluids, respectively, such that the density and viscosity of the fluids obey the following functional dependence with temperature. 
\begin{eqnarray}
\rho_A &=&1200 -0.9 T, \\
\rho_B &=&2504 -2.84 T,\\
\mu_A &=&{\rm exp}(-10.17+1643/T), \\
\mu_B &=&{\rm exp}(-11.76+1540/T).
\end{eqnarray}
Note that these relationships are different from the general formulation given in Section \ref{sec:formulation}. The thermal conductivity and the heat capacity of the dispersed and continuous phases are kept as constants, such that $\kappa_A=0.13389$ W/mK,  ${c_p}_A=1778.2$ J/kg K and $\kappa_B=0.0063$ W/mK,  ${c_p}_A=1047$ J/kg K.

\begin{figure}[h]
\centering
\includegraphics[width=0.6\textwidth]{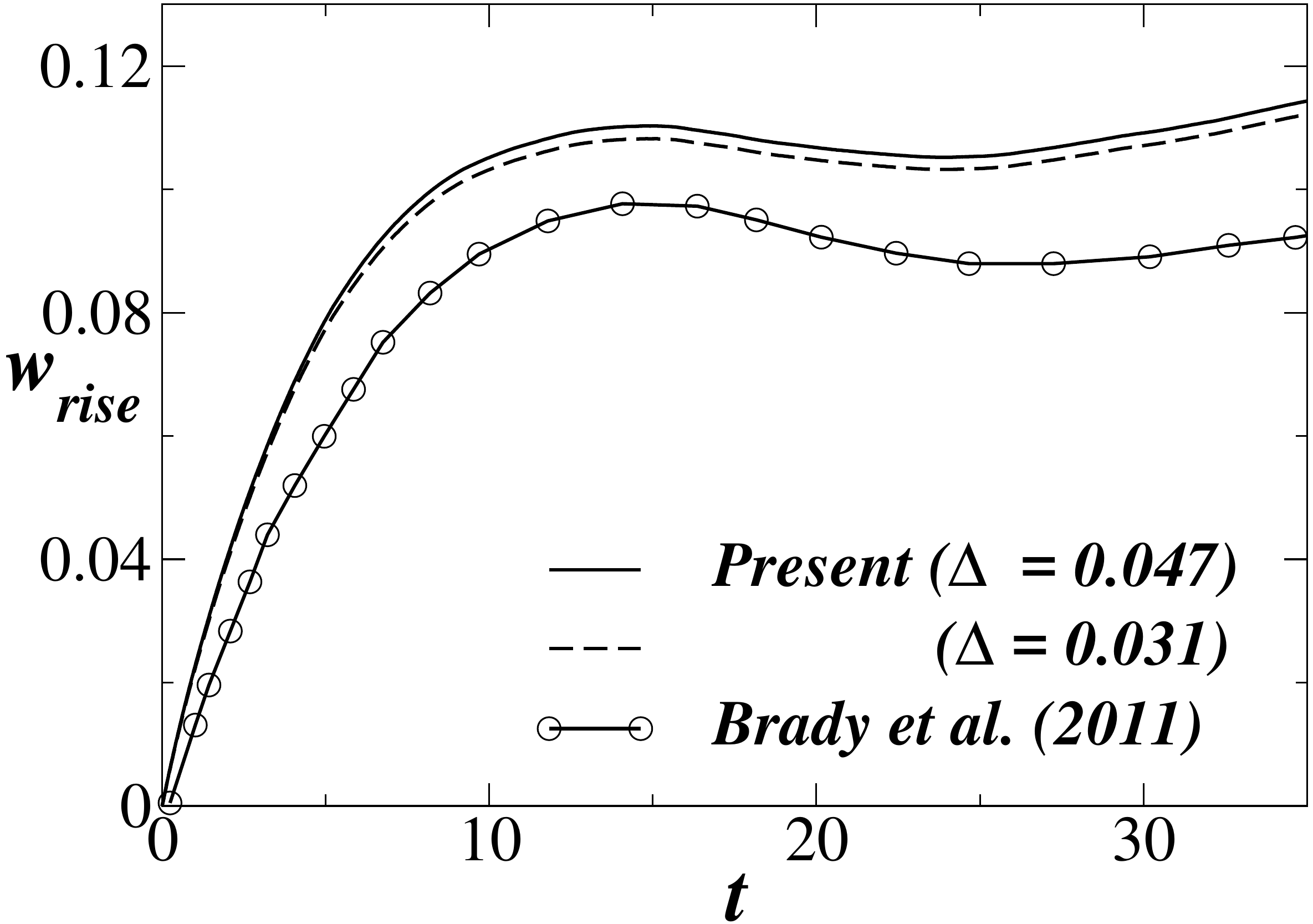}
\caption{Temporal variations of the rise velocity of a bubble normalised with $\beta \gamma R/\mu_A$.}
\label{brady}
\end{figure}

Brady {\it et al.} \cite{Brady2011} used the radius of the bubble, $R$, $\beta \gamma R/\mu_A$ and $\gamma R$ as the length, velocity and temperature scales, based on which the corresponding dimensionless numbers are $Re=17.79$, $Ma=1723$ and $Ca=0.0275$. As such a high value of $Ma$ leads to a thin thermal boundary layer, which is difficult to resolve numerically, they used $Ma=86$ instead of $Ma=1723$ in their numerical simulation. Thus, we also used $Ma=86$ in our numerical simulation for this case.

\begin{figure}[h]
\centering
\includegraphics[width=0.8\textwidth]{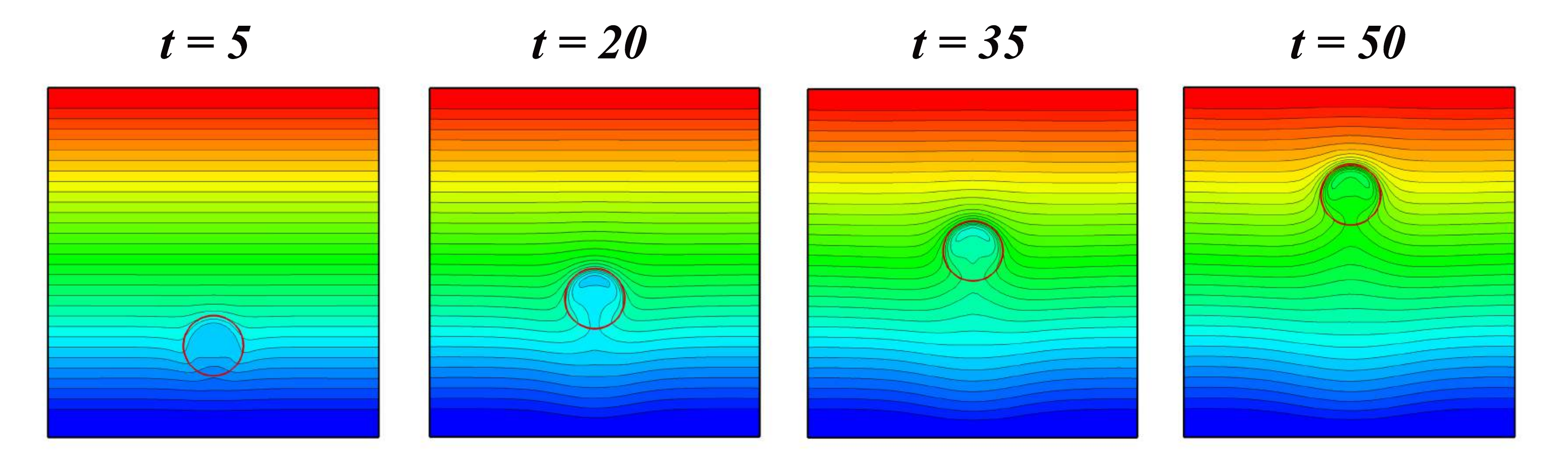}
\caption{Isotherms plotted at $t=5$, 20, 35 and 50. The shape of the droplet is shown by red circle. The parameters are the same as those used to generate Fig. \ref{brady}.}
\label{brady_isotherns}
\end{figure}

We have conducted the numerical simulation for the same set of parameters as used by Brady {\it et al.} \cite{Brady2011}.
The temporal variations of dimensional rise velocity of the bubble normalised with $\beta \gamma R/\mu_A$ are shown in Fig. \ref{brady}. Adoptive grid refinement is used in our simulations with the size of the smallest grids equal to 0.047 and 0.031. It can be seen that the maximum difference between the rise velocity obtained using these two grids is about 1\%. The solid line with circle symbols represents the result of Brady {\it et al.} \cite{Brady2011} and our result is shown by the solid black line. It can be seen that our simulation give slightly higher rise velocity as compared to that obtained by Brady {\it et al.} \cite{Brady2011}. Now if we inspect Fig. \ref{hermann}, we observe that their simulations under-predict the theoretical value of Young {\it et al.} \cite{Young1959}. Also, the computational domain considered in the present study is slightly different for the one used by Brady {\it et al.} \cite{Brady2011}. Thus, we may attribute the difference observed in Fig. \ref{brady} to these effects. However, the isotherm contours obtained using the same parameters as those used to generate Fig. \ref{brady} agree well with those given in Brady {\it et al.} \cite{Brady2011} (see Fig. \ref{brady_isotherns}). In order to gain more confidence for simulations associated with high Marongani numbers, we have considered another test case studied by Liu {\it et al.} \cite{liu2012} in Section \ref{liu12}.

\subsubsection{Comparison with Nas \& Tryggvason \cite{Nas2003}, Seric {\it et al.} \cite{seric2017} and Ma \& Bothe \cite{Ma2011}}
\label{seric}

Another test case is considered to study the thermocapillary migration of a droplet in another fluid in the microgravity condition. Both 2D and 3D simulations are performed and migration velocity of the droplet is compared with that of the previous studies \cite{Nas2003,seric2017,Ma2011}. This test case was first considered by Nas \& Tryggvason \cite{Nas2003} and subsequently used in the recent studies to validate their numerical solvers. In our study, square and cubic computational domains with $H=4R$ are used for two and 3D simulations, respectively. As considered by Nas \& Tryggvason \cite{Nas2003}, a droplet of initial radius $R=1.44$ mm with fluid properties $\rho_B=500$ kg m$^{-3}$, $\mu_B=0.024$ Pa$\cdot$s, $\kappa_B= 2.4 \times 10^{-6}$ Wm$^{-1}$K$^{-1}$ and ${c_p}_B=10^{-4}$ J Kg$^{-1}$K$^{-1}$ is kept at the centre of the computational domain. The ratio of the fluid properties of the ambient fluid with those of the drop is 2. $\sigma_0=10^{-2}$ Nm$^{-1}$ and $\gamma=2 \times 10^{-3}$ Nm$^{-1}$K$^{-1}$. In order to compare with the result of Nas \& Tryggvason \cite{Nas2003}, the following dimensionless numbers were used by Ma \& Bothe \cite{Ma2011} and Seric {\it et al.} \cite{seric2017}:
\begin{equation}
Re={\rho_B R U_r \over \mu_B}, \quad Ma={\rho_B {c_p}_B R U_r \over \kappa_B}, \quad {\rm and} \quad Ca={\mu_B U_r \over \sigma_0}. 
\label{Ma}
\end{equation}
Here, the reference velocity, $U_r = \gamma \nabla T R / \mu_B$ and $\sigma_0$ is the surface tension at the reference temperature, $T_0$. In their formulation
\begin{equation}
\sigma = 1 -Ca (T-T_0),
\label{Ma1}
\end{equation}
wherein $T_0=290$ K and the temperature difference between the top and bottom walls, $\nabla T$ = 200 Km$^{-1}$. The above mentioned physical properties gives $Re=Ma=0.72$ and $Ca=0.0576$ in the dimensionless formulation (i.e. Eq. (\ref{Ma})). 

\begin{figure}[h]
\centering
(a) \hspace{7.5cm} (b) \\
\includegraphics[width=0.45\textwidth]{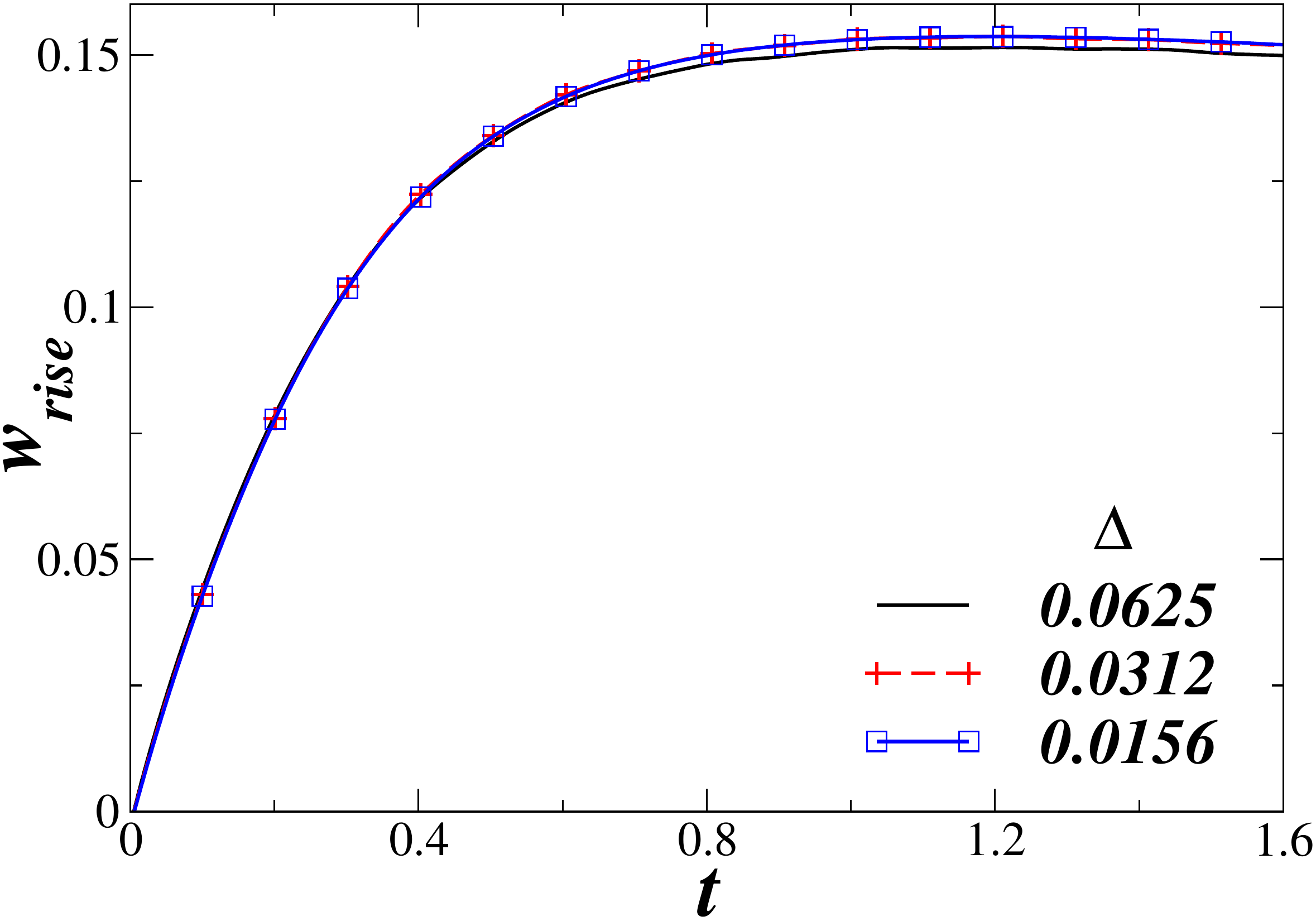} \hspace{2mm} \includegraphics[width=0.45\textwidth]{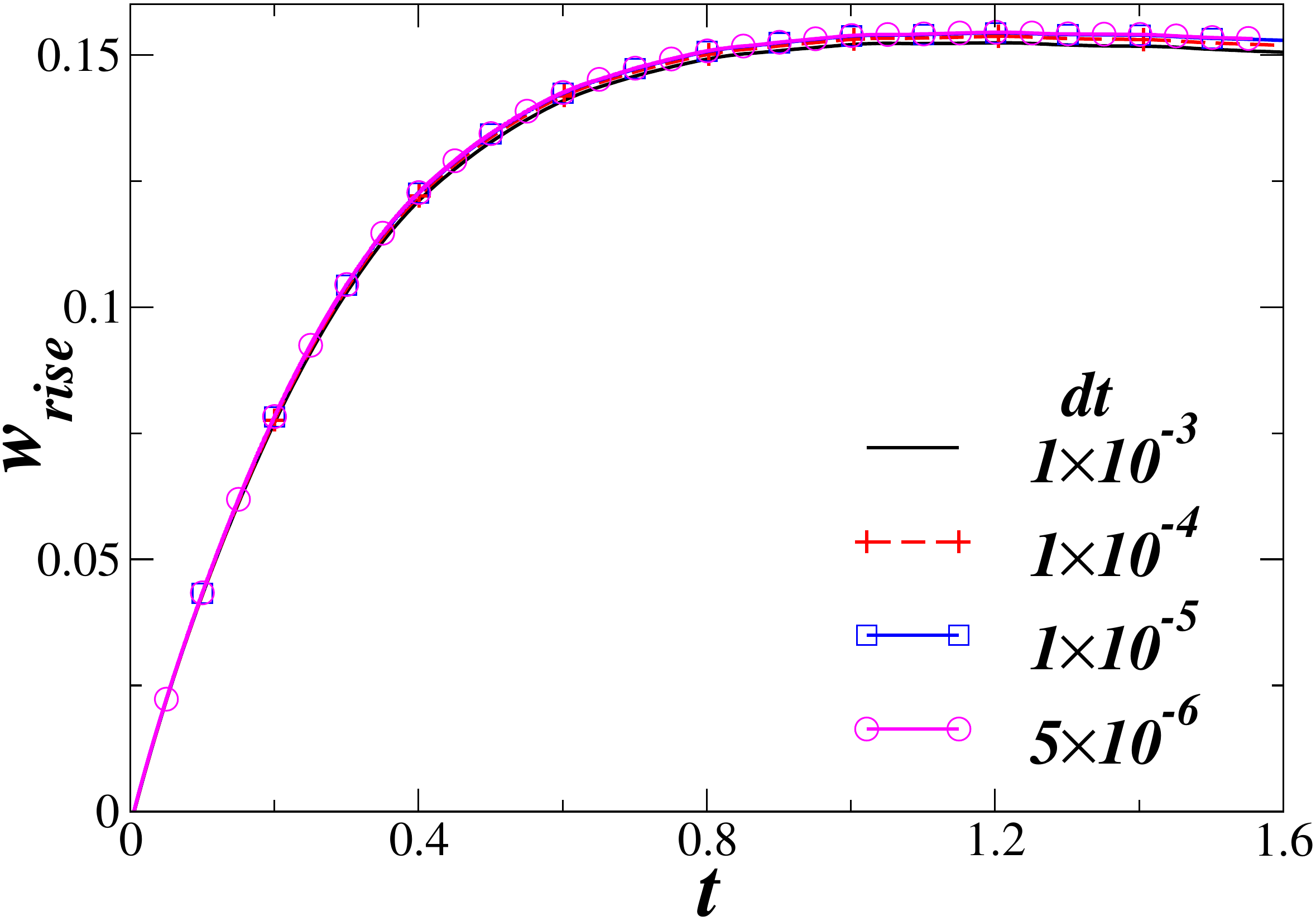}
\caption{(a) Grid convergence test (with time step, $dt = 1 \times 10^{-4}$) and (b) effect of time step, $dt$ (with grid size, $\Delta = 0.0078$) on drop migration velocity for $Re = Ma = 0.72$ and $Ca = 0.0576$.}
\label{grid}
\end{figure}

We have conducted numerical simulations for the same set of parameters as discussed above. In our numerical simulations, no-slip and no-penetration boundary conditions are used at the top and bottom walls, whereas the Neumann boundary conditions for the velocity components and temperature are used at the size boundaries. First a grid convergence test is conducted using uniform grid sizes, $\Delta=0.0625$, 0.0312 and 0.0156 as shown in Fig. \ref{grid}(a). It can be seen that there is negligible difference between the results obtained using $\Delta=0.0312$ and 0.0156, whereas less than 1.5 \% error is observed between the results obtained using $\Delta=0.0312$ and 0.0625. The error is defined as
$$\left (1 - {w_{rise}|_{\Delta=0.0312} \over w_{rise}|_{\Delta=0.0625}} \right ) \times 100.$$
Similarly, we also study the effect of time step on the rise velocity of the droplet for this case and found converged results even for $dt=10^{-3}$. 

\begin{figure}[h]
\centering
\includegraphics[width=0.6\textwidth]{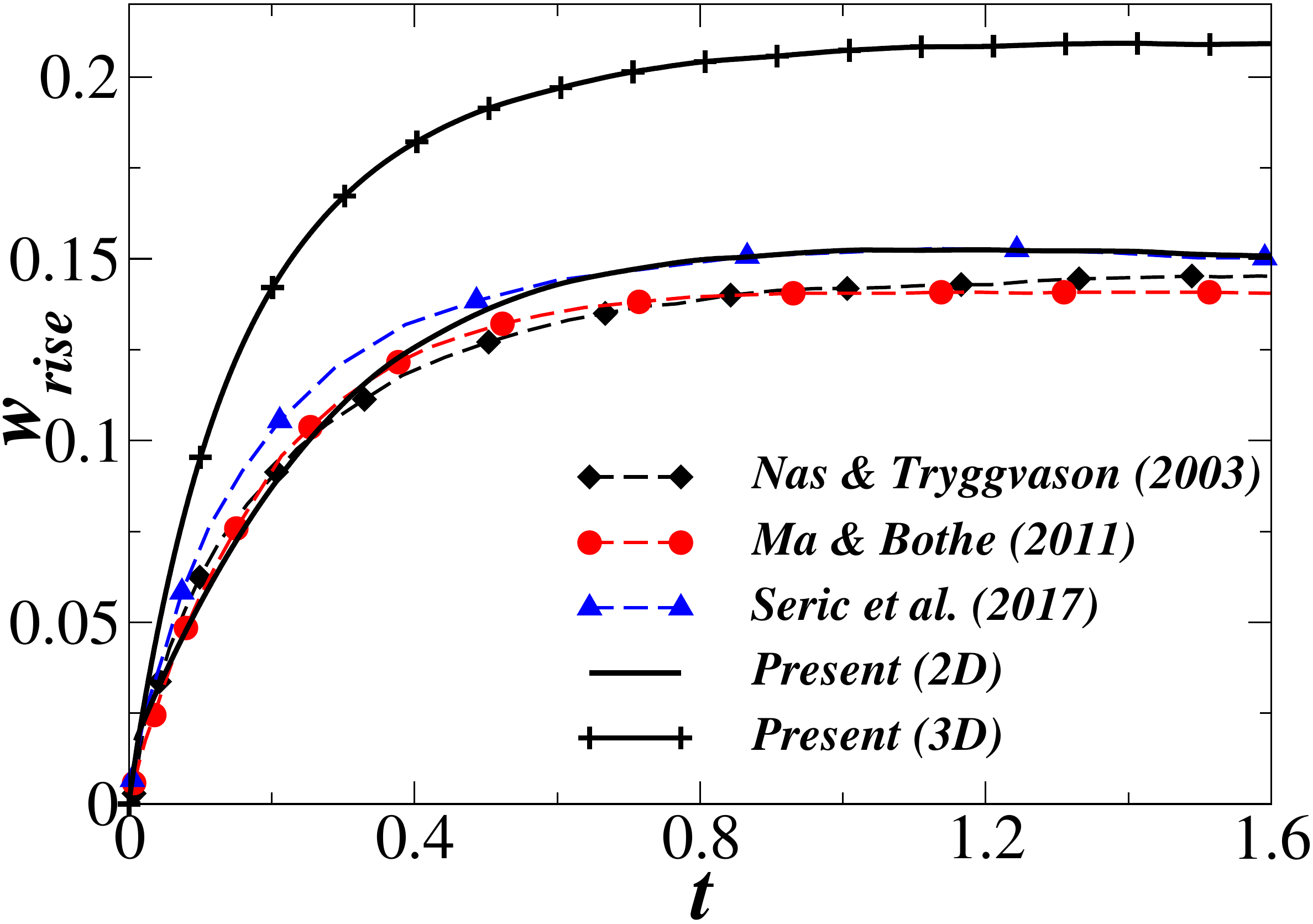}
\caption{Drop migration velocity for $Re = Ma = 0.72$ and $Ca = 0.0576$.}
\label{afkami}
\end{figure}

In Fig. \ref{afkami}, the temporal variations of the rise velocity of the droplet normalised with $U_r$ obtained from our two and 3D simulations are shown, along with the results of the previous studies \cite{Nas2003,seric2017,Ma2011}. Note that the previous studies presented only 2D results. Fig. \ref{afkami} reveals that the terminal rise velocity obtained from our 2D simulation agrees well that of Seric {\it et al.} \cite{seric2017}. However, in the accelerating regime ($t<0.4$), our result is closed to that of Nas \& Tryggvason \cite{Nas2003} and Ma \& Bothe \cite{Ma2011}. Our 3D simulation predict a much higher rise velocity as compare to the 2D case. This is expected as the three-dimensional droplet experiences a stronger Marangoni force as compared to the two-dimensional droplet. 

\subsubsection{Comparison with Liu {\it et al.} \cite{liu2012}}
\label{liu12}
\ks{In this section, the objective is to compare the thermocapillary migration of a bubble, particularly for a high Marangoni number, obtained from the present simulation with that of Liu {\it et al.} \cite{liu2012}. The study of Liu {\it et al.} \cite{liu2012} was 3D, but we conducted both axisymmetric and 3D simulations for the same case. The list of parameters used by Liu {\it et al.} \cite{liu2012} are given in Table I. The properties the fluids, i.e. the values of kinematic viscosity, density and thermal conductivity of fluid $A$ and fluid $B$ are assumed to be the same. $k_A=k_B=0.002$ is used to obtain the Marangoni number $Ma$ equals to 100. The top and bottom walls are maintained at temperature 0 and 24. As they have used a lattice Boltzmann method, all these parameters were in the lattice units. In the dimensionless formulation, using $R$ as the length scale, ${\beta \gamma R / \mu_A}$ as the velocity scale and $T_{ref}$ as the temperature scale, we get $Re=1$, $Ca=0.1$ and $\Gamma=0.13333$ (refer Table 1 for the list of the dimensionless numbers used in the present study). The flow dynamics is simulated in a computational domain of size $15 \times 15 \times 15$ using the smallest grid size $(\Delta)$ equals to $\approx$ 0.06. A similar grid was considered by Liu {\it et al.} \cite{liu2012}; however, the width and breadth of the channel were 7.5. Like in their case, no-slip and no-penetration boundary conditions are used at the top and bottom walls, and periodic boundary conditions for the velocity components and temperature are used at the side boundaries. Also, the present simulations are conducted using adaptive grid refinement, which provides finer grid near the interfacial region (near bubble) and slightly courser grid in the outer region. In our axisymmetric case, half of the computational domain in the $x$ direction is used and the dynamics is assumed to be symmetrical about $x=0$.}

\begin{table}[h]
\centering
\caption{Parameters considered by Liu {\it et al.} \cite{liu2012} (in lattice units) and the corresponding dimensionless parameters in our study.}
\begin{tabular}{|c|c|c|c|c|c|c|c|c|c|}
\hline
\multicolumn{10}{|c|}{\bf Parameters considered by Liu {\it et al.} \cite{liu2012} (in lattice units)} \\ \hline
R  & $T_{ref}$ & $\sigma_{0}$ & $\beta$    & $\rho_A$  &  $\nu_A$ & $k_A$ & $U_{YGB}$  & $V$ & $\gamma$\\ \hline
16 & 12 & $2.5 \times10^{-2}$ & $-1.5625 \times10^{-3}$    & 1  &  0.2 & 0.002 & $1.667 \times 10^{-3}$  &0.0125 & 0.1 \\ \hline
\multicolumn{10}{|c|}{\bf Dimensionless parameters used in the present study} \\ \hline
\multicolumn{3}{|c|}{$Re (\equiv VR / \nu_A)$} & \multicolumn{2}{|c|}{$Ca (\equiv \rho_A V \nu_A / \sigma_{0})$} &  \multicolumn{2}{|c|}{$\Gamma (\equiv \gamma R / T_{ref})$} & \multicolumn{2}{|c|}{$Ma (\equiv R V \nu_A / k_A)$} & \multicolumn{1}{|c|}{$ M (\equiv T_{ref} \beta / \sigma_0)$}\\ \hline
\multicolumn{3}{|c|}{1} & \multicolumn{2}{|c|}{0.1} &  \multicolumn{2}{|c|}{0.1333} & \multicolumn{2}{|c|}{100} & \multicolumn{1}{|c|}{0.75} \\ \hline
 \end{tabular}
\label{budget}
\end{table}

\begin{figure}[h]
\centering
\includegraphics[width=0.6\textwidth]{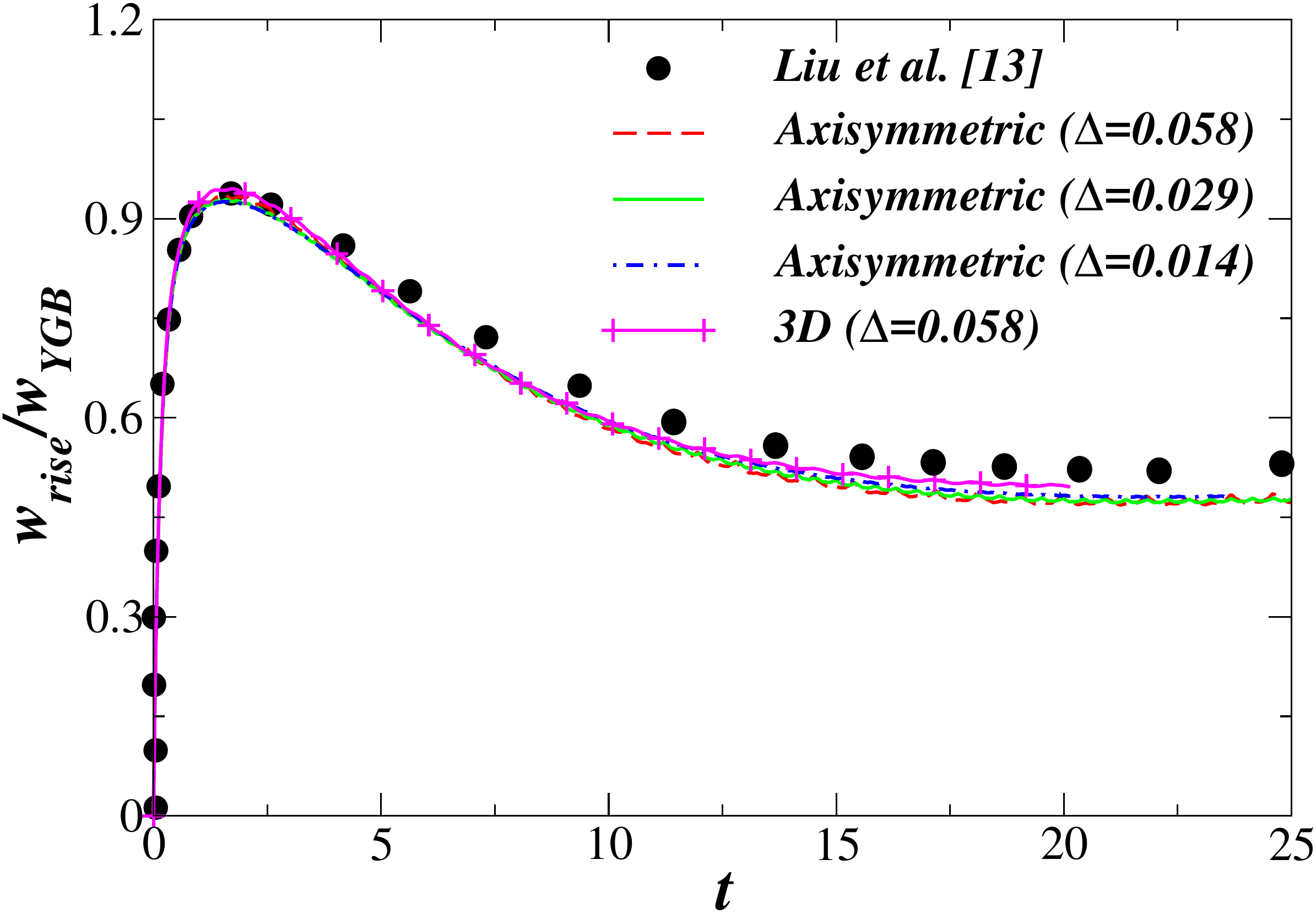}
\caption{The temporal variations of the normalised velocity of the bubble for $Ma=100$. The rest of the parameters are $Re=1$, $Ca=0.1$ and $\Gamma \approx 0.133$.}
\label{liucomp}
\end{figure}

\ks{In Fig. \ref{liucomp}, the temporal variations of the bubble rise velocity normalised with the theoretical result of Young {\it et al.} \cite{Young1959} (for $Ma=0$) are plotted. A grid convergence test is also conducted by performing simulations using 
$\Delta = 0.058$, 0.029 and 0.014 in our axisymmetric simulations. It can be seen that the results are indistinguishable confirming that the grid convergence test. In view of this, a 3D simulation is performed using $\Delta = 0.058$, and the 3D result is shown by the solid line with plus symbols. It can be seen that the maximum rise velocity and the dynamics at early times compare well again Liu {\it et al.} \cite{liu2012}. However, our 3D simulation slightly under-predicts the result of Liu {\it et al.} \cite{liu2012} at later times. }

\subsection{Migration of an air bubble inside a liquid medium under the action of thermocapillary force and buoyancy}

So far thermocapillary migration of a bubble has been investigated in the reduced gravity condition. In this section, the dynamics of an air bubble under the simultaneous action of thermocapillary force and buoyancy is studied. The formulation used in this section is exactly the same as the one described in Section \ref{sec:formulation}. In this case, an initially spherical bubble of radius $R$ is placed at $z_i=10R$ in a computational domain of $H=20R$. The bubble rise dynamics is investigated for different values of $M$, i.e dimensionless $-d \sigma/dT$. For $M>0$, the force due to the surface tension gradient and buoyancy act in the same direction (in the positive $z$ direction). On the other hand, for $M<0$, the force due to the surface tension gradient acts in the negative $z$ direction, while buoyancy acts in the positive $z$ direction. 

\begin{figure}[h]
\centering
\includegraphics[width=0.6\textwidth]{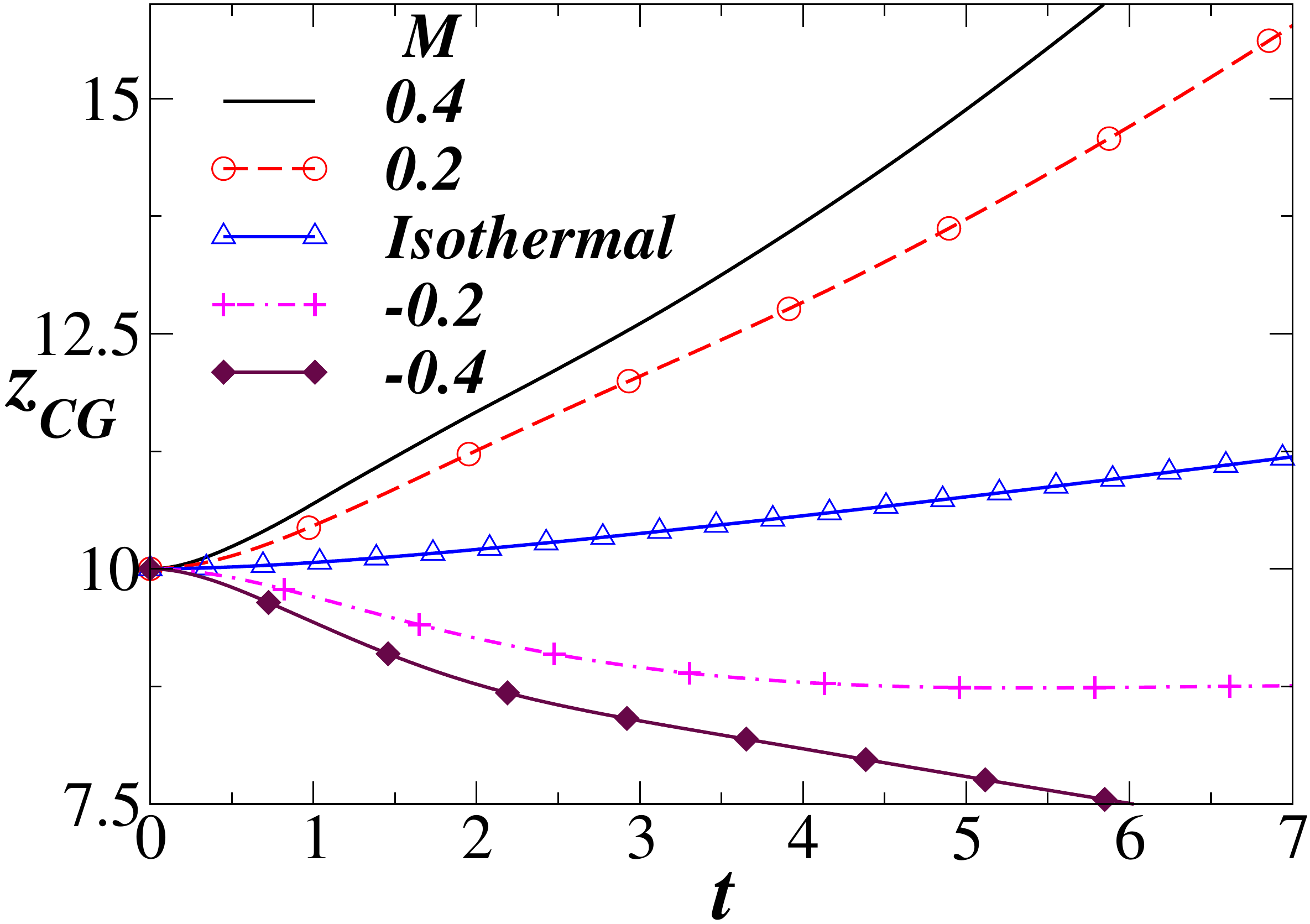}
\caption{The temporal variations $z_{CG}$ of an air bubble rising in a liquid medium for different values of $M$. The rest of the parameters are $Re=10$, $Ca=0.01$, $Fr=10$, $Pr=7$, $\alpha_r=0.04$, $\mu_r=10^{-2}$, $\rho_r=10^{-3}$ and $\Gamma=0.1$.}
\label{zcg}
\end{figure}

\begin{figure}[h]
\centering
\hspace{5mm} (a) \hspace{28mm} (b) \hspace{28mm} (c) \hspace{28mm} (d) \\
\includegraphics[width=0.9\textwidth]{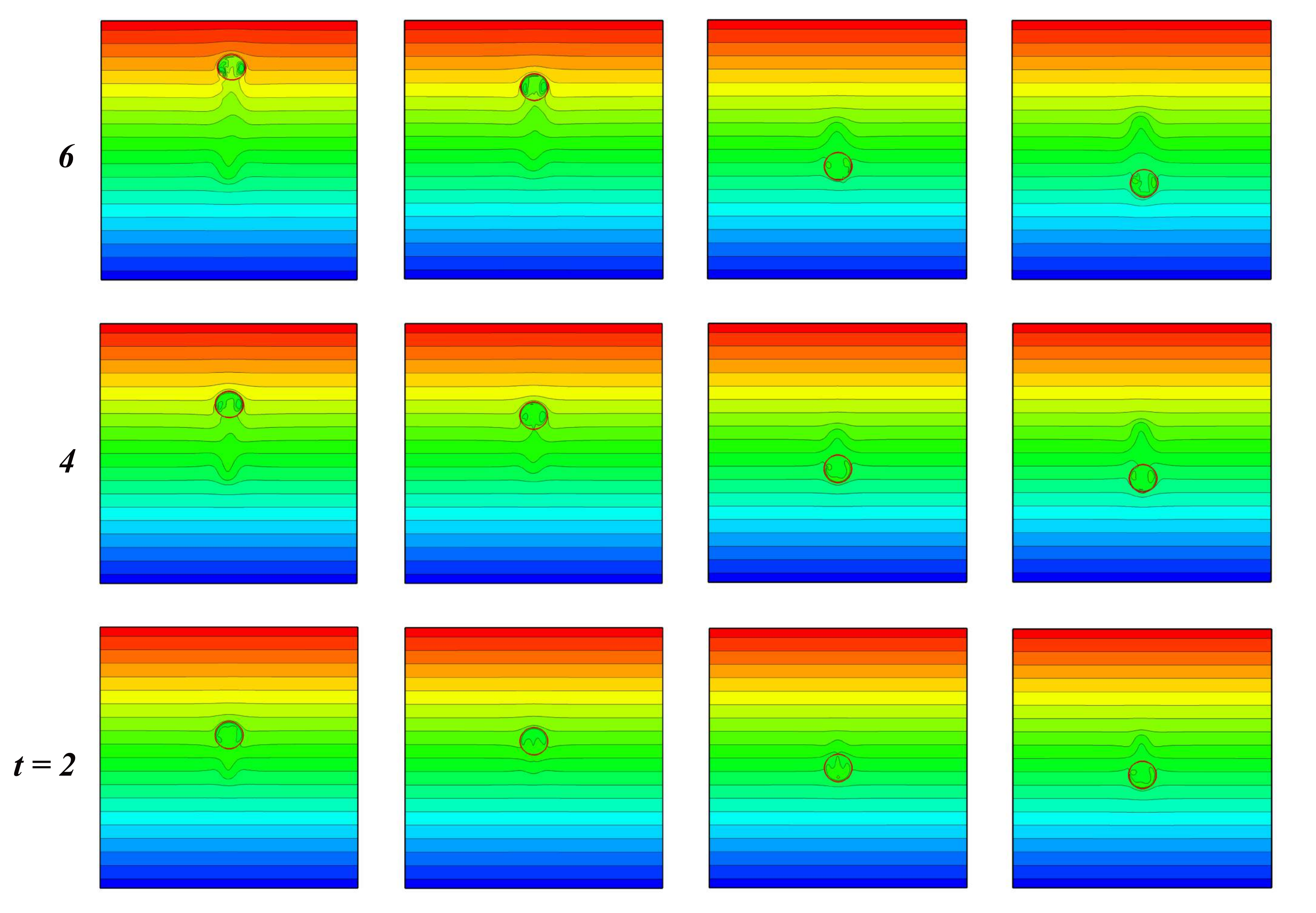}
\caption{Isotherms plotted at $t=2$, 4 and 6 for (a) $M=0.4$, (b) $M=0.2$, (c) $M=-0.2$ and (d) $M=-0.4$. The parameters are the same as those used to generate Fig. \ref{zcg}.}
\label{isotherms}
\end{figure}

\begin{figure}[h]
\centering
\includegraphics[width=0.9\textwidth]{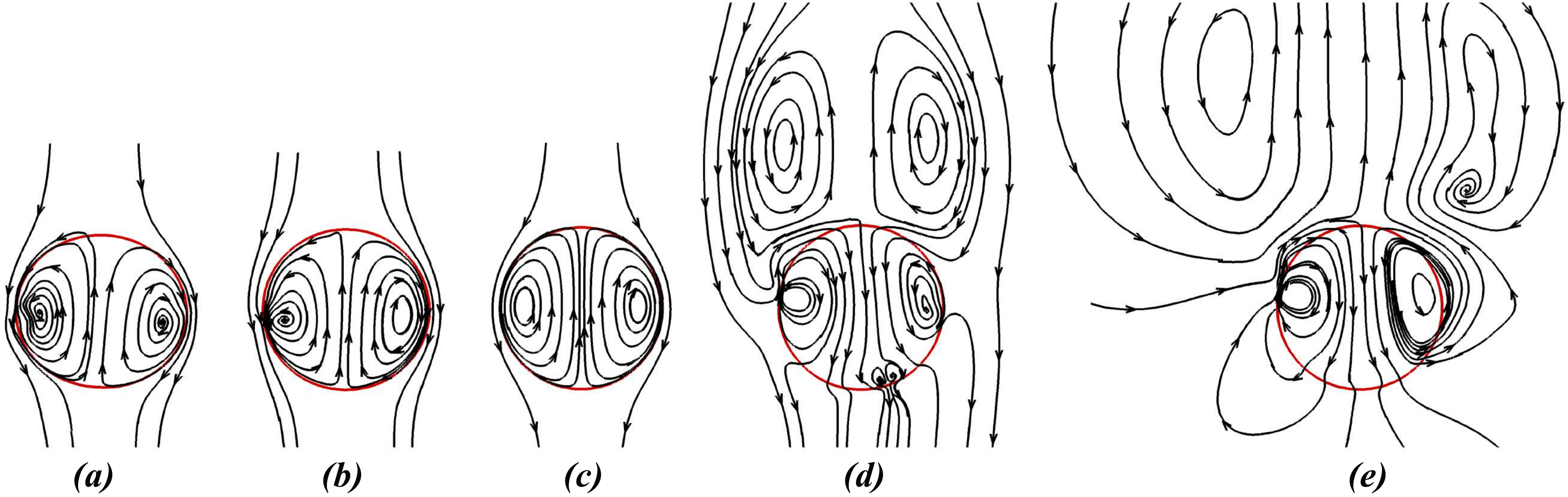} \\
\caption{Streamlines at $t=5$ for (a) $M=0.4$, (b) $M=0.2$, (c) isothermal, (d) $M=-0.2$ and (e) $M=-0.4$. The rest of parameters are the same as those used to generate Fig. \ref{zcg}.}
\label{stream2}
\end{figure}

In Fig. \ref{zcg}, the temporal variations of the centre of gravity of the air bubble, $z_{CG}$ are plotted for different values of $M$. The rise dynamics is also compared with the isothermal case. The result obtained for an isothermal system is shown by line with open triangle symbols. In this case, it can be seen that the bubble rises and attains a terminal velocity (slope of the line is constant at later time). For $M>0$, as expected, increasing the value of $M$ increases the velocity of the bubble rising in the upward direction. One interesting phenomena is observed for $M=-0.2$. As discussed above, for $M<0$, the surface tension force acts in the {direction opposite to that of} the buoyancy force. This in turn opposes the upward motion of the bubble due to buoyancy. For $M=-0.2$ after going through a decelerating phase, the bubble gets arrested at $z \approx 8.85$. This implies that the surface tension force which is acting in the negative $z$ direction for $M=-0.2$ balances the buoyancy force. Based on a creeping flow analysis, the condition of bubble arrest, as derived by Young {\it et al.} \cite{Young1959} (also see \cite{Leal1992}), is given by
\begin{equation}
M_{cr} = -{2 \over 3} {Re Ca \over Fr \Gamma}.
\end{equation}
For the parameters considered in Fig. \ref{zcg}, the critical value of $M$ at which a bubble gets arrested in the creeping flow is $-0.0667$. However, in our numerical simulations, we note that the bubble gets arrested for $M = -0.2$. This shows the importance of a three-dimensional non-linear flow analysis in case of Marangoni flows even for spherical bubbles. For $M=-0.4$, it can be seen that the surface tension force dominates the flow and the bubble migrates in the downward direction. The isotherms in the $x$-$z$ plane passing through the centre of gravity of the bubble are plotted at different times for $M=0.4$, 0.2, -0.2 and -0.4 in Fig. \ref{isotherms}. It can be seen that the isotherms inside the bubble become asymmetrical due to the Marangoni flow. The shape of the bubble {(shown by red line)} reveals that for $M=0.4$ the bubble deforms to an oblate shape at later times ($t \ge 6$), whereas it remains spherical for the rest of the $M$ values considered. This is due to the fact that for large positive $M$ (say $M=0.4$) the resultant inertial force due to the surface tension gradient and the buoyancy, which act in the same direction in this case, dominates the flow as compared to the surface tension {and viscous forces}, unlike the other cases where the surface tension force wins, which keeps the bubble in a spherical shape. The streamlines patterns in the the $x$-$z$ plane are shown at $t=5$ for different values of $M$ and isothermal system in Fig. \ref{stream2}. As expected, Hadamard \cite{hadamard1911} type steady flow field is observed for the isothermal case. For positive values of $M$ the flow gets distracted slightly and also becomes unsteady. For negative values of $M$, a big recirculation zone appears at the top as the bubble migrates in the negative $z$ direction. The size of this recirculation zone increases with increasing the negative value of $M$.

\section{Concluding remarks}
\label{sec:conc}
In this work, we demonstrate a novel way to handle surface tension gradient driven flows in the VoF framework. An open source Navier-Stokes solver, {\it Basilisk} is used, and the present formulation is implemented within this solver to study thermocapillary flows. A characteristic problem, where thermal Marangoni stresses play a significant role, namely, the thermocapillary migration of drops and bubbles in a surrounding medium is considered. By performing several validation exercises, we have shown that our solver is very robust {and accurate} to investigate interfacial flows with variable surface tension. In such class of problems, calculating surface tension force tangential and normal to the interface separating the fluids is very challenging. In order to overcome the numerical difficulties, most of the studies use numerical tricks, such as smearing the surface tension force about the interface. Due to this their results always under-predict the theoretical prediction. On the other hand, the present method {employs a second order accurate height-function-like method to compute the surface tension gradient along the interface}. We have shown that our results predict the theoretical terminal velocity of a droplet migrating due to an imposed temperature gradient derived by Young {\it et al.} \cite{Young1959}. Also, most of the previous computational studies, investigate thermocapillary flows in the microgravity condition (i.e. by neglecting gravity). However, Merritt {\it et al.} \cite{Merritt1993} demonstrated that the systems which experience both buoyancy and thermocapillary forces simultaneously exhibit complex flow structures and the intuition developed by including the forces separately is not good enough. Thus, we investigate the rise dynamics of an air bubble inside a liquid medium under the action of both thermocapillary and buoyancy forces. Finally, we would like to remark that the present numerical solver could be used to study interfacial flows with surface tension gradients {(not limited to thermocapillary flows)} accurately.

\noindent{\bf Acknowledgement:} K. C. S. thanks Indian National Science Academy for their financial support. M. K. T. also thanks the Department of Science \& Technology, India (Project number: DST/EES/2015037) for financial support. We also thank Mounika Balla for the help in plotting some of the results presented here. 

\end{document}